%% file: main.tex
\newif\ifdouble
\newif\ifsingle
\newif\ifchange

\documentclass[sigconf]{acmart}

\newif\ifdouble

\ifdouble
  
\else
   
\fi

\usepackage{listings}
\usepackage{xcolor} 

\newcommand{\system}[1]{Generative Lecture}

\usepackage{xcolor}
 
\colorlet{RED}{red}
\usepackage[normalem]{ulem}
\newcommand\redsout{\bgroup\markoverwith{\textcolor{red}{\rule[0.3ex]{2pt}{1.2pt}}}\ULon}

\usepackage{tabularx}
\usepackage{subcaption}

\usepackage{dblfloatfix} 
\usepackage{booktabs} 
\usepackage{arydshln}
\usepackage{multirow}
\usepackage{makecell}
\acmConference[arXiv]{}{December, 2025}

\begin{document}

\title{\system{}: Making Lecture Videos Interactive with LLMs and AI Clone Instructors}

\author{Hye-Young Jo}
\orcid{0000-0003-3847-3420}
\affiliation{%
  \institution{University of Colorado Boulder}
  \city{Boulder}
  \country{USA}}
\email{hye-young.jo@colorado.edu}

\author{Ada Yi Zhao}
\orcid{} 
\affiliation{%
  \institution{University of Colorado Boulder}
  \city{Boulder}
  \country{USA}}
\email{ada.zhao@colorado.edu}

\author{Xiaoan Liu}
\orcid{} 
\affiliation{%
  \institution{University of Colorado Boulder}
  \city{Boulder}
  \country{USA}}
\email{xiaoan.liu@colorado.edu}

\author{Ryo Suzuki}
\orcid{0000-0003-3294-9555} 
\affiliation{%
  \institution{University of Colorado Boulder}
  \city{Boulder}
  \country{USA}}
\email{ryo.suzuki@colorado.edu}

\renewcommand{\shortauthors}{Jo, et al.}
\renewcommand{\shorttitle}{\system{}: Making Lecture Videos Interactive with LLMs and AI Clone Instructors}

\input{sections/0-abstract}

\begin{teaserfigure}
\includegraphics[width=\textwidth]{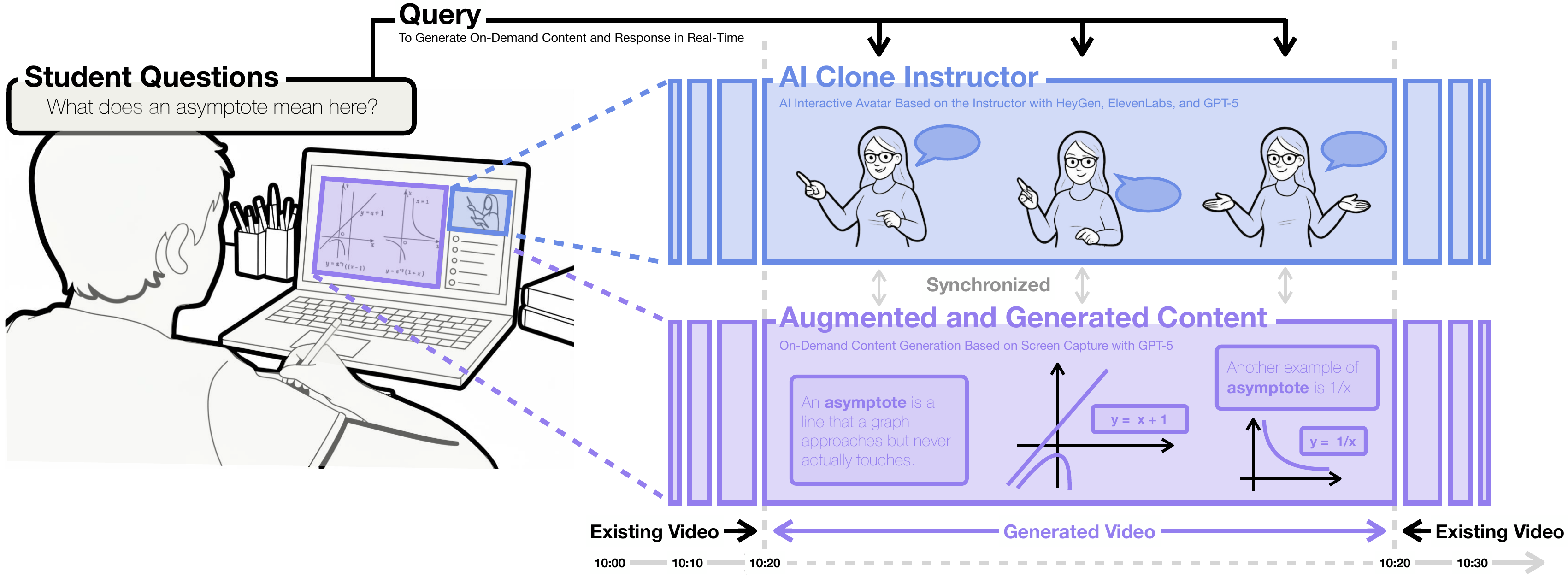}
\caption{Generative Lecture is a concept to transform existing lecture videos into interactive, personalized learning experiences with an AI clone instructor and generated content. When students ask questions, the system generates and inserts new material into the lecture slides, providing embedded responses such as on-demand clarification, augmented visuals, and adaptive examples alongside the original video. Then, the AI clone instructor, powered by HeyGen, ElevenLabs, and GPT-5, explains the content synchronously as if the instructor in the video were directly responding to the student’s question.}
\Description{}
\label{fig:teaser}
\end{teaserfigure}

\maketitle

\input{sections/1-introduction}
\input{sections/2-related-work}

\input{sections/3-formative-study}

\input{sections/4-system}

\input{sections/5-evaluation}
\input{sections/6-discussion}
\input{sections/7-conclusion}

\ifdouble
  \balance
\fi
\bibliographystyle{ACM-Reference-Format}
\bibliography{references}

\clearpage

\input{sections/appendix}

\end{document}

%% file: sections/0-abstract.tex
\begin{abstract}
We introduce Generative Lecture, a concept that makes existing lecture videos interactive through generative AI and AI clone instructors. By leveraging interactive avatars powered by HeyGen, ElevenLabs, and GPT-5, we embed an AI instructor into the video and augment the video content in response to students’ questions. This allows students to personalize the lecture material, directly ask questions in the video, and receive tailored explanations generated and delivered by the AI-cloned instructor. From a design elicitation study (N=8), we identified four goals that guided the development of eight system features: 1) on-demand clarification, 2) enhanced visuals, 3) interactive example, 4) personalized explanation, 5) adaptive quiz, 6) study summary, 7) automatic highlight, and 8) adaptive break. We then conducted a user study (N=12) to evaluate the usability and effectiveness of the system and collected expert feedback (N=5). The results suggest that our system enables effective two-way communication and supports personalized learning.
\end{abstract}

\ccsdesc[500]{Human-centered computing}

\keywords{Interactive Videos, Personalized Learning, Large Language Models, Generative AI, Virtual Agent, AI Clone}

%% file: sections/1-introduction.tex
\section{Introduction}
We envision that the future of lecture videos will be interactive and bi-directional. Currently, lecture video content on platforms like Coursera and Khan Academy remains the same for every viewer---once published, regardless of the viewer's needs and context, every single viewer watches exactly the same content. Thus, current video-watching experiences are essentially passive and one-directional. What if existing lecture videos could dynamically change and respond based on user needs and questions? In this way, viewers could truly interact with videos, just like in-person lectures, so that video content can continuously adjust and evolve over time, personalizing for each user's context and needs.

In this paper, we introduce \system{}, a concept to make existing lecture videos interactive and bi-directional by leveraging generative AI and AI clone instructors. By modifying, inserting, and augmenting video materials on the fly and in real-time with generative AI, students can directly ask questions and receive tailored explanations delivered by the AI-clone instructor, along with content augmentation (Figure~\ref{fig:teaser}). This concept and design are motivated by a formative interview and design elicitation study with eight university students. The study results suggest a strong need for seamless integration of AI and video content. For example, students mentioned that they sometimes continue watching without fully understanding the material, and that pausing the videos to ask ChatGPT disrupts their focus, making it harder to stay on track. On the other hand, they value lecture videos for their structured scaffolding, which provides context and guidance that standalone tools like ChatGPT cannot offer. In the study, participants also designed possible interactions with lecture videos based on their current practice and prior experiences interacting with tutors.

Based on our formative study, we identified four learning goals that motivated eight desired interactions: 
1) \textit{On-Demand Clarification}: the instructor can respond to students’ questions both verbally and visually, 
2) \textit{Enhanced Visual}: the system augments slides with external figures to simplify complex concepts, 
3) \textit{Interactive Diagram}: the system generates an interactive diagram that students can manipulate to deepen understanding, 
4) \textit{Personalized Explanation}: the system provides analogies and examples tailored to the student’s background or interests, 
5) \textit{Adaptive Quiz}: the system delivers quizzes with adjustable difficulty to assess understanding at a self-paced pace,
6) \textit{Study Summary}: the system automatically captures slides and student questions, enabling students to revisit content and interact with the AI avatar for review, 
7) \textit{Automatic Highlight}: the system synchronizes the instructor’s speech with slides, automatically annotating and highlighting important points, and 
8) \textit{Adaptive Break}: the system schedules short breaks and generates light but educational content relevant to the lecture and the student’s interests. 
We implemented these features in a proof-of-concept system that integrates response generation (GPT-5), voice synthesis (ElevenLabs), instructor avatars (HeyGen), slide recognition (Gemini), and content generation (HTML Canvas).

We evaluated our system from both instructor and student perspectives. On the student side, we conducted a within-subjects study with twelve participants, comparing our system against a baseline video with interactive quizzes and unrestricted access to online search (e.g., ChatGPT and other web sources). We measured usability, engagement, and its effectiveness in two video lectures: 1) deep learning in computer science, and 2) particle physics. From the instructor's perspective, we conducted expert interviews with five university instructors to evaluate the system’s potential to support content creation workflows, enhance student engagement, examine the acceptability of cloned avatars, and identify areas for improvement. Results show that the system reduced students’ frustration and increased their engagement during the learning process. Instructors acknowledged the system’s potential to reduce instructional workload and elicit honest feedback on their lectures, but expressed concerns about the reliability of AI-generated visuals and the trustworthiness of cloned avatars speaking on their behalf.

In summary, our contributions include
\begin{enumerate}
\item The \system{} concept, comprising four pedagogical design goals and eight interaction designs, derived from a formative design elicitation study with university students (N=8).
\item An implementation of the system, an interactive lecture video platform that uses an LLM-based AI-clone instructor to generate personalized content embedded within the original lecture video.
\item Two user studies---1) a within-subject study with students (N=12) and 2) an expert interview with instructors (N=5)---that investigate the impact on learners' engagement and effectiveness from both student and teacher perspectives.
\end{enumerate}

\begin{figure}[h]
\centering
\includegraphics[width=\columnwidth]{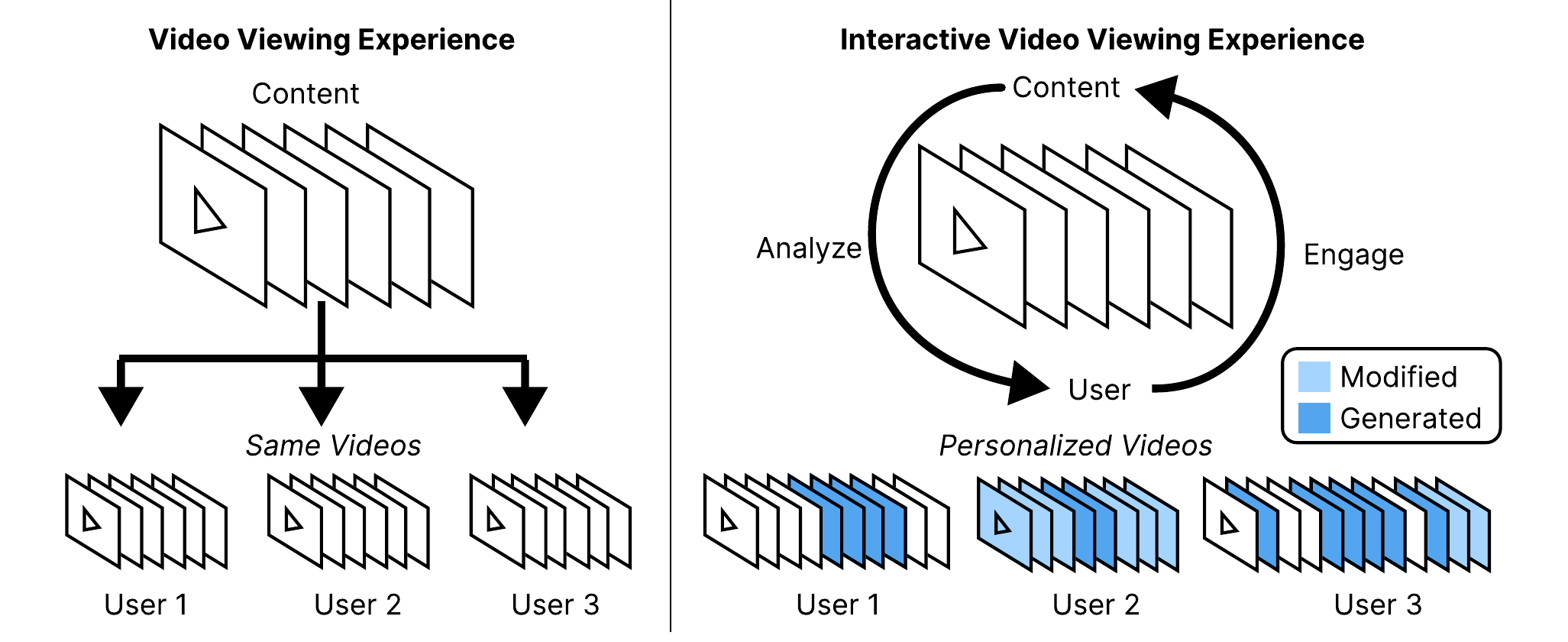}
\caption{A concept of interactive and bi-directional videos. In traditional videos (left), the content is predetermined, offering the same experience to all viewers. In contrast, interactive videos (right) adapt dynamically to each viewer's needs by modifying existing content, adding new elements, and seamlessly integrating them into the video to create a personalized and responsive experience. While this concept is generalizable to any type of video, in this paper, we specifically focus on lecture videos to make them interactive.}
\label{fig:concept}
\Description{}
\end{figure}

%% file: sections/2-related-work.tex
\section{Related Work}
In this section, we review prior work on interactive video interfaces and the use of generative AI in education.

\subsection{Interactive Video Interfaces}
While some interactive video interfaces focus on the authoring experience~\cite{wang2024lave, leake2017computational}, and automatic creation of instructional videos from documents, web pages, or demonstrations~\cite{chi2012mixt, chi2021howtocut, chi2022doc2video, chi2020url2video, truong2021automatic, zhong2021helpviz}, our review centers on those that enhance the viewing experience through branching narratives~\cite{netflix2018bandersnatch, de2018video}, adaptive playback~\cite{pavel2015sceneskim, clarke2020reactive}, and annotation functionality~\cite{zhou2019magic}.

\subsubsection{Branching Narrative}
Branching narratives allow viewers to influence the storyline by making real-time choices. This interactive storytelling ranges from a simple selection of pre-recorded footage using clickable hot spots~\cite{elnahla2020black} to the real-time composition of multiple foreground and background elements~\cite{de2018video}. Hyper-video platforms, such as Annoto, H5P, EdPuzzle, and Kaltura, offer branching features that enable viewers to choose narrative paths and take interactive quizzes. In education, digital storytelling has been shown to enhance engagement~\cite{quah2022systematic, kim2015rimes}. Platforms like H5P have been found to enhance understanding in asynchronous learning~\cite{rama2022enhanced, mir2021investigation}. 

\subsubsection{Adaptive Playback}
Adaptive playback enhances video interaction through within-video navigation, synchronization with external media, and spatial adaptation. Some systems improve navigation by linking video with keywords, enabling search through transcripts and summaries or automatic playback adjustments~\cite{pavel2015sceneskim, chang2021rubyslippers, truong2021automatic}. Other systems synchronize video with external documents, such as PDFs or slides, ensuring relevant content is displayed alongside the video without requiring manual searching~\cite{kim2023papeos, murakami2024swapvid}. Some interfaces adapt video positioning based on user posture or gaze, either by repositioning the video itself to maintain visibility~\cite{shin2020body} or modifying subtitle placement within the region of interest~\cite{kurzhals2020view}.

\subsubsection{Annotation Overlay}
The annotation feature makes video consumption more social and participatory by allowing users to add content directly to videos. Text-based annotations, such as Danmaku systems, overlay real-time comments onto videos, enabling viewers to share reactions and create a shared viewing experience~\cite{chen2017watching, wu2018danmaku}. In education, they help students engage with lecture videos by highlighting key points to take notes or ask questions~\cite{lin2018exploratory, mirriahi2016uncovering, yang2024aqua}, or linking additional materials for collective learning~\cite{kim2021hyperbutton}. Sketch-based annotations provide a visual layer of interaction, enabling creative collaboration in live streams~\cite{lu2021streamsketch} and facilitating strategy discussions in gameplay analysis~\cite{riegler2014videojot}. Closer to our overlay approach, VideoSticker~\cite{cao_videosticker} introduces sticker-like, in-situ annotations that attach explanations to specific regions of a video, supporting situated learning. Whereas VideoSticker emphasizes viewer-authored, object-anchored notes, our system \textit{generates} new visuals and explanations on demand and can \textit{initiate} interactions, shifting from static overlays to bi-directional, adaptive content.

\addvspace{0.3cm}
\noindent
In summary, interactive video has been widely studied, but existing systems offer only uni-directional, static interaction, where user input affects the presentation but not the content itself. We propose a new form of \textit{\textbf{bi-directional}} interactive video, in which the video acts as a living entity, initiating interaction and incorporating \textit{\textbf{dynamically evolving content}} that adapts to user input.

\subsection{Generative AI in Education}
Generative AI enables scalable yet personalized learning~\cite{almadhoob2024quizwiz} by providing tailored explanations~\cite{abolnejadian2024leveraging, chen2024gptutor, leong2024putting} and simulating learning environments~\cite{blau2021writing, liu2024classmeta} for students, while also assisting instructors in preparing lecture materials~\cite{zeghouani2024examining, han2024teachers, fan2024lessonplanner}. This section reviews LLM-based pedagogical agents, including text-based chatbots and AI avatars, as well as AI-generated educational content in various forms, such as text, images, and videos.

\subsubsection{LLM-Based Pedagogical Agents}
Unlike early rule-based chatbots, such as AutoTutor~\cite{graesser2005autotutor}, which follow predefined dialogue structures with limited adaptability, LLM-powered chatbots generate customized explanations and quizzes on demand~\cite{jin2024teachtune, seo2021impact}, assist with complex tasks such as code generation and debugging~\cite{kazemitabaar2024codeaid}, and simulate diverse human behaviors and personal traits through fine-tuning~\cite{jin2024teachtune, calo2024towards}. As a result, AI chatbots like ChatGPT are already widely used in self-directed and online learning~\cite{rajala2023call}, and  LLM-based agents serve multiple roles in education, including instructors or tutors~\cite{jin2024teachtune, khokhar2022modifying}, peers~\cite{jin2024teachtune, wang2025generative, liu2024classmeta}, and idealized self-representations that support motivation~\cite{zheng2023self}. 

AI avatars enhance chatbot capabilities by incorporating speech, facial expressions, and gestures, creating a more interactive and engaging learning experience. Developed using deepfake technology and LLMs~\cite{pataranutaporn2021ai}, these avatars simulate human-like communication, making digital learning environments more immersive~\cite{khokhar2022modifying, johnson2000animated, pataranutaporn2022ai, pataranutaporn2023living, liu2024classmeta}. However, some research suggests that AI tutors or instructors may lack the social presence and perceived credibility of their human counterparts~\cite{lee2024teachers, xu2024recorded, lim2024potential}. Consequently, AI tutors are often viewed as a complement to human educators rather than a replacement~\cite{lee2024teachers, kim2020my, han2024teachers, bahroun2023transforming}.

\subsubsection{Educational Content Generation}
Generative AI has enabled the adaptation of educational materials to better align with students’ learning needs. AI-generated text dynamically adjusted explanations based on learners’ background knowledge, interests, and learning preferences~\cite{abolnejadian2024leveraging, chen2024gptutor, pesovski2024generative}. In language education, AI-generated personalized stories and contextual examples to enhance comprehension~\cite{leong2024putting, peng2023storyfier}, while in programming, AI-generated tailored hints encouraged problem-solving by guiding students without providing direct solutions~\cite{kazemitabaar2024codeaid}. Instructors also benefited from AI-assisted tools that streamlined lesson planning~\cite{fan2024lessonplanner} and provided automatic content variations, including different instructional styles~\cite{pesovski2024generative}, as well as assessments and feedback~\cite{fang2024edulive}.

Beyond text, AI-generated visuals improved conceptual understanding for students while reducing the effort required for content creation. For example, ChemGenX~\cite{abbas2024chemgenx} facilitated chemistry education by generating molecular structures tailored to specific learning objectives. AI-generated comics enhanced language learning by incorporating contextual storytelling~\cite{panchal2024lingocomics}. Additionally, AI-generated videos expanded instructional content by transforming static slides into dynamic presentations~\cite{dao2021ai, xu2024recorded} or generating video explanations directly from text~\cite{weerakoon2024enhancing}. While these tools enhanced accessibility and engagement, they required domain-specific fine-tuning to ensure accuracy and pedagogical effectiveness.

\addvspace{0.3cm}
\noindent
Building upon this landscape of research, we explore how AI-generated avatars and personalized materials can be integrated into existing lecture videos. By \textit{\textbf{embedding}} LLM-based avatars and AI-generated content into the existing video, we enhance engagement and adaptability in video-based learning while preserving the structure of human instruction.

%% file: sections/3-formative-study.tex
\section{Reimagining Interactive Lecture Videos: Formative Interviews and Design Elicitation Study}
To explore how lecture videos could become more interactive and personalized, we conducted formative interviews and design elicitation studies with eight university students. Through these interviews and design sessions, we examined participants’ current study practices, identified challenges in studying with lecture videos, and elicited interaction design ideas to address their challenges.

\subsection{Participants}
We recruited eight participants (one undergraduate, three master's students, and four PhD students) from diverse departments with STEM backgrounds (Table~\ref{tab:participant_profiles}). Participants were recruited via the university student email list and snowball sampling, and they were screened for prior experience of studying with lecture videos. Participants had viewed lecture videos covering various STEM subjects: Mathematics (P1-P8), Programming (P1, P2, P7, P8), Physics (P2, P4, P6, P8), Computer Science (P3, P4, P7, P8), Electrical Engineering (P1, P4, P5), and Aerospace Engineering (P2).

\begin{table}[!htbp]
  \centering
  \begin{tabular}{@{}llll@{}}
    \toprule
        & Program   & Major                     & User-Selected Video Lecture                                                            \\ 
    \midrule
    P1  & MS        & HCI                       & \href{https://youtu.be/XO8u0Y75FRk?si=BPj1WN1i6OiB3pWO}{Processing}                       \\
    P2  & PhD       & Aerospace Eng.            & \href{https://youtu.be/KKAD-OOOHxg?si=cgg5EKc5GDgG3v4L}{Organic Chemistry}                \\
    P3  & MS        & HCI                       & \href{https://youtu.be/Ye018rCVvOo?si=OR0jAZxAKqwv68EE}{Machine Learning}                 \\
    P4  & PhD       & HCI                       & \href{https://youtu.be/pHmRp2eGETk?si=RNqYMLgnmQHTP9Bf}{Quantum Computing}                \\
    P5  & Undergrad & Biomedical Eng.           & \href{https://www.coursera.org/specializations/computational-neuroscience}{Neuroscience}  \\
    P6  & MS        & Physics                   & \href{https://youtu.be/ojPQN86BW9o?si=AB-HGLsaJzOeBVkZ}{Particle Physics}                 \\
    P7  & PhD       & CS                        & \href{https://www.coursera.org/learn/network-systems-foundations}{Network System}         \\
    P8  & PhD       & CS                        & \href{https://youtu.be/qxXmpkkgO2o?si=V74Fv9V7qOoSakEO}{Nonlinear Dynamics}               \\
    \bottomrule
  \end{tabular}
  \caption{Participant Profiles and Their Study Materials}
  \label{tab:participant_profiles}
\end{table}

\subsection{Protocol}
The study had two phases: \textbf{\textit{1) Formative Interview:}} investigating students' learning behaviors and strategies in video-based learning through observation and contextual interviews, and \textbf{\textit{2) Design Elicitation Study:}} envisioning personalized, interactive lecture videos through a design activity. The study was conducted online (Zoom) and lasted 90 minutes per participant.

\subsubsection{Formative Interviews}
In the first phase, we aimed to understand learning behaviors through a survey and a contextual interview, which lasted about 60 minutes. Each participant completed two tasks: a survey and a demo. First, they filled out a Google survey about their video-based learning experiences, and the experimenter asked follow-up questions to gain a deeper understanding of their responses. Next, each participant watched two lecture videos: one they had previously studied and another provided by the experimenter. They demonstrated their learning process using a think-aloud protocol. Table~\ref{tab:participant_profiles} presents the list of user-selected videos (embedded as hyperlinks). For the given video, we selected \textit{Matrix Examples – Geometric Transformations and Selectors}~\footnote{\url{https://youtu.be/jXN50fuRSqE}} (06:02) from \textit{Stanford University's Applied Linear Algebra series}~\footnote{\url{https://youtube.com/playlist?list=PLoROMvodv4rMz-WbFQtNUsUElIh2cPmN9}}
, as it introduces fundamental data processing concepts relevant to STEM students across various disciplines. Through Zoom screen sharing, participants demonstrated their learning setup and procedure for the first video, describing how they approached and resolved challenges when they did not understand the content. For the second video, they demonstrated their study process using any tools they normally relied on, such as online search engines, LLM-powered assistants, or physical note-taking methods. After observing, the experimenter asked follow-up questions to clarify participants’ intentions behind their study behaviors and their opinions on the differences between learning through lecture videos and in-person lectures, particularly in terms of interaction.

\subsubsection{Design Elicitation}
In the second stage, we aimed to envision personalized interactive lecture videos through a 30-minute design activity. The experimenter introduced the concept of \textbf{\textit{interactive videos}} (Figure~\ref{fig:concept}), which feature \textit{bi-directional interaction} and \textit{real-time content modification}. Participants were then asked to redesign lecture videos to make them more interactive and personalized. To facilitate this, the experimenter pre-captured frames from the given lecture video, converted the narration into speech bubbles, and prepared them as editable slides on an online whiteboard (Miro). Using text boxes, diagrams, and drawings, participants modified these frames to personalize the content and enhance engagement. The experimenter then asked them to describe how and when they would like these modifications to be triggered and assisted with the drawing process. To guide their design process, we provided examples, such as an instructor asking students if they were paying attention or a video automatically updating slides based on a user’s question. Next, participants applied the same approach to their user-selected video. They identified moments where they had struggled to digest the content, captured key frames, and designed interactive features that could have helped them at those points.

\subsubsection{Post-Study Analysis}
For analysis, we transcribed all interview recordings and compiled participants' redesigned lecture video slides on a shared whiteboard for analysis. Using thematic coding~\cite{braun2006using}, we extracted key quotes from the transcripts to identify participants' perceived issues in the lecture videos, both in terms of content and presentation, as well as their motivations for making modifications. These insights were then mapped to the corresponding design artifacts to connect participants' reasoning with their design choices. We categorized the design artifacts based on the types of modifications participants made, their interaction preferences, and the purposes behind these changes. The first author developed an initial set of labels, which was iteratively refined through discussions among all authors until no new themes emerged, ensuring consistency and comprehensiveness.

\subsection{Results} 
\begin{figure*}[h]
    \centering
    \includegraphics[width=\textwidth]{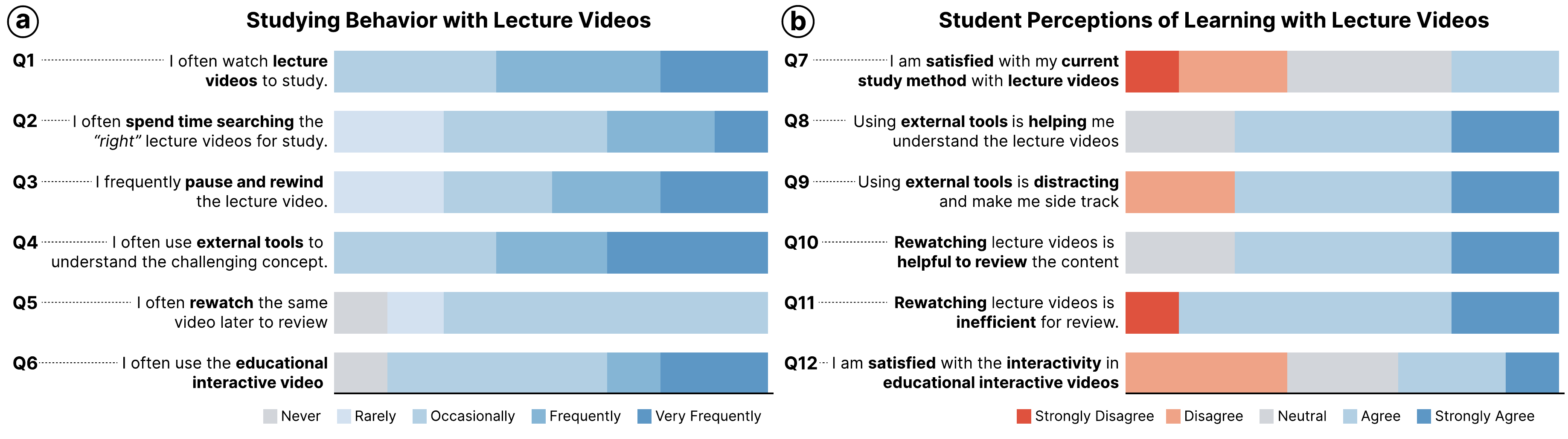}
    \caption{Survey result.}
    \label{fig:survey-result}
    \Description{}
\end{figure*}

Figure~\ref{fig:survey-result} presents survey results on how students use lecture videos for studying and their views on their effectiveness. Overall, the lecture videos discussed during the study fell into two categories: (1) \textbf{\textit{university lecture recordings}}, which were generally long, complex, and slide-based but valued for being “well-structured” (P4, P7), “in-depth” (P7), and “conversational” (P5); and (2) \textbf{\textit{general educational videos}}, which were shorter, visually engaging, and often included animations or sketches. Students primarily used university lecture recordings as their main resource and general educational videos as supplementary materials, particularly for hands-on practice (P1) or to understand basic concepts (P3).

\begin{figure*}[h]
\centering
\includegraphics[width=\textwidth]{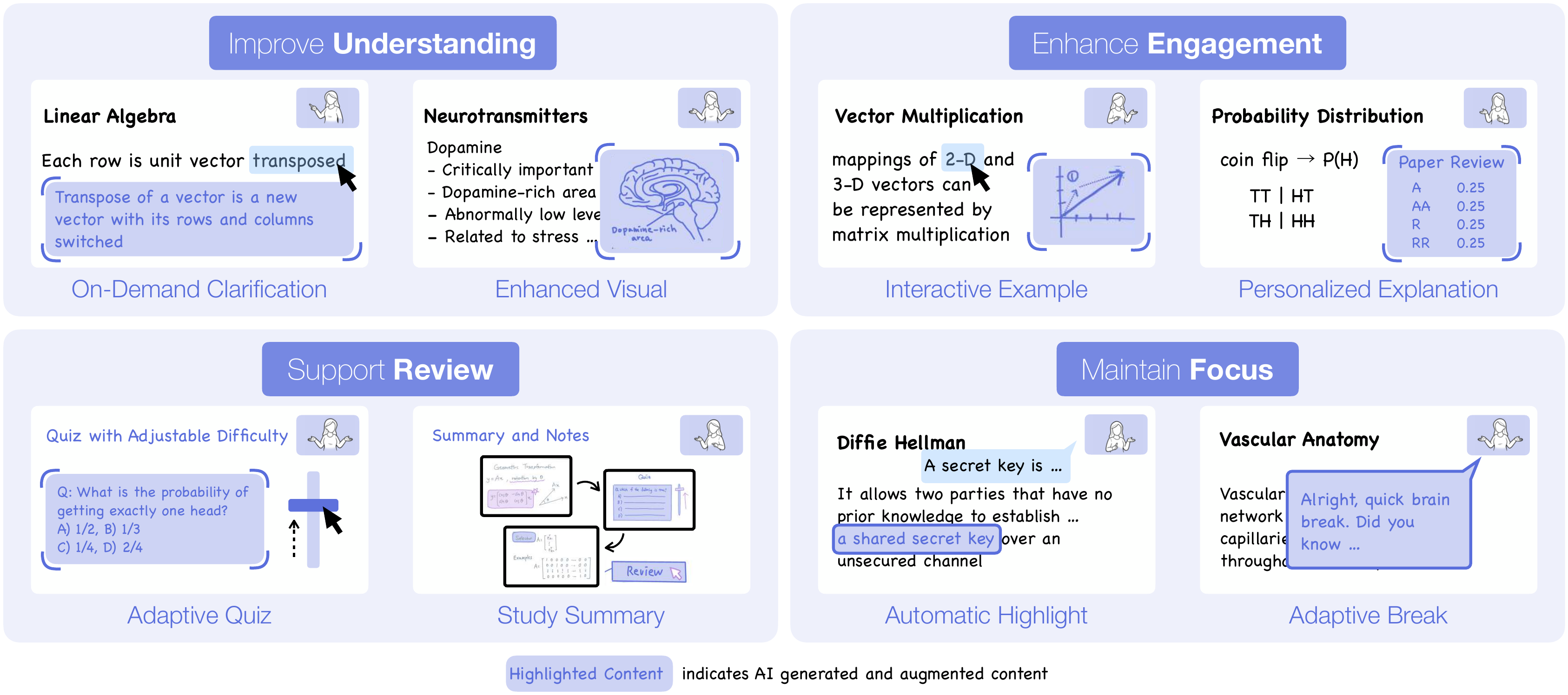}  
\caption{Four themes and eight interaction designs identified from participants' lecture video redesign artifacts.}
\label{fig:design-space}
\Description{}
\end{figure*}

Through the design elicitation, participants suggested a total of 88 redesign ideas for the videos---55 based on the provided videos and 33 from videos of their own selection. Each participant contributed between 7 and 15 suggestions ($M = 11$, $SD = 3.02$). After excluding single-case ideas, we grouped the remaining ideas into four learning challenge categories related to studying with lecture videos, and distilled recurring desired interactions into eight design artifacts (Figure~\ref{fig:design-space}).

\subsection{Theme 1: Improve Understanding}
\subsubsection{Challenge: Filling the Knowledge Gap or Missing Information}
Students spent considerable time selecting lecture videos to find one that best suited their background knowledge and preferred teaching styles. They searched between 5 and 120 minutes for the \textit{right} video ($M = 31.25$, $SD = 39.98$), considering factors such as well-structured content (P1, P7), appropriate difficulty levels (P2, P3), and visually rich materials (P2, P4, P8). However, unsurprisingly, no single video fully met their needs. Students often had to fill in missing background knowledge while skipping over familiar concepts. To supplement gaps, they turned to external resources such as ChatGPT (P1–P4, P6–P8), web searches (P1–P4, P7, P8), textbooks (P2, P5), and peers (P5) (Q4). While most students (6 out of 8) found these tools helpful (Q8), frequently switching between multiple resources often led to distractions (Q9).

\subsubsection{Elicited Design: On-Demand Clarification}
Instead of searching in a separate browser, participants wanted to directly select content and receive quick clarifications to maintain immersion. Many sought quick definitions (P3, P4, P7, P8) or translations (P5, P7) of new terminology. Some preferred step-by-step explanations when instructors skipped mathematical calculations (P4, P7), while others wanted alternative explanations or problem-solving approaches, particularly in programming (P6, P8). These preferences were shaped by prior experiences with web searches and large language models (LLMs), such as ChatGPT and DeepSeek. When using LLMs, participants expected concise responses, with P7 and P8 emphasizing the need for cross-validation to ensure the trustworthiness of the results.

\subsubsection{Elicited Design: Enhanced Visual}
Participants preferred visually expressive lecture content and wanted to replace or supplement text with visuals to better follow the lecture. Participants (P5, P7) found text-heavy slides overwhelming and favored visually appealing content. Many (P1–P5) mentioned searching for relevant images or videos to better understand applications discussed in lecture videos or to explore unfamiliar concepts. To address this, participants suggested embedding relevant visuals, including 2D images (P2, P4, P5, P7), 3D models (P2, P5), animations (P2, P8), and videos (P1, P3).

\subsection{Theme 2: Enhance Engagement}
\subsubsection{Challenge: Maintaining Motivation}
Because recorded lecture videos are inherently one-way communication, students found it difficult to maintain motivation. To stay engaged while watching, some students rewound specific sections (P1–P3, P5–P7), while others paused the video to take notes (P2, P3, P4) (Q3). Students also noted that they preferred lecture videos where instructors used fun, real-world examples (P5) and provided hands-on exercises they could work on after watching (P7).

\subsubsection{Elicited Design: Interactive Example}
Participants (P3, P7, P8) found numerical values more intuitive than abstract symbols (e.g., \( x, y, \theta \)) for understanding concepts. Some participants (P1, P7) wanted to manipulate numbers or components in figures and observe the results dynamically.

\subsubsection{Elicited Design: Personalized Explanation}
Participants wanted more relevant and practical examples, noting that lecture video examples often felt “flat” (P4) and “not intuitive” (P7). Many (P1, P3, P4, P6, P8) preferred examples tied to their own field of study or daily life. For instance, instead of using a coin flip to explain probability distribution, one participant (P3) suggested using paper acceptance rates.

\subsection{Theme 3: Support Review}
\subsubsection{Challenge: Reviewing and Retaining Knowledge}
Most students (7 out of 8) had experience with interactive educational videos (Q6) and found embedded quizzes useful for checking their understanding (Q12). However, some (P4, P7, P8) felt the quizzes were too easy and less engaging compared to tutoring (P3, P5). Since the quizzes often repeated video content, they did not encourage “deeper thinking” (P7). To better assess their knowledge, some students created their own questions (P2) or discussed concepts with friends (P5). For review, students found rewatching lecture videos helpful (Q10), but most avoided rewatching entire lectures due to time constraints (Q5). Instead, they preferred skipping to specific sections, though this was difficult when videos lacked timestamps or clear chapters (Q12). Some students took notes while watching to minimize the need for rewatching, while many relied on online searches or LLMs to quickly retrieve forgotten information.

\subsubsection{Elicited Design: Adaptive Quiz}
To check their understanding while watching lecture videos, adaptive quizzes were suggested. For embedded quizzes available in many educational platforms, participants wanted the ability to adjust difficulty levels, as they found existing quizzes too easy (P1, P2, P6). In addition, one participant (P8) suggested automatic navigation based on quiz results, while another (P3) proposed skipping familiar content based on a pre-quiz.

\subsubsection{Elicited Design: Study Summary}
To further support reviewing lecture content after some time had passed, participants wanted interactive summaries following the videos. Participants (P1–P4, P6) suggested the ability to revisit personalized videos whenever needed and proposed a summary note with links to specific timestamps to reduce search time. Some (P2, P4, P6) also wanted to save their questions and notes for future reference.

\subsection{Theme 4: Maintain Focus}
\subsubsection{Challenge: Staying Focused During Passive Viewing Experience}
Even after selecting the appropriate lecture videos, maintaining focus remained challenging (Q7). Presentation issues such as text-heavy slides (P7, P8), unclear explanations (P1, P4), and monotonous delivery (P5) made it difficult to stay engaged. To counter this, some students took notes while watching (P2, P5) to stay focused, while others switched to different videos when they lost interest (P4, P6).

\subsubsection{Elicited Design: Automatic Highlight}
To reduce the cognitive effort of mapping lecture content with the instructor's narration, several participants wanted a visual indicator of where the instructor was explaining. They suggested highlighting (P2, P4, P7, P8) or tracing text with a mouse pointer (P2) or a hand (P3, P5, P6). Some participants missed parts of the content because they prioritized the instructor’s verbal explanations, but found it easier to follow when the instructor sketched over the slides. Meanwhile, some (P2, P4, P5) recommended breaking information-dense slides into smaller chunks and creating multiple slides to improve clarity.

\subsubsection{Elicited Design: Adaptive Break}
Beyond content modification, participants envisioned the instructor or the video itself initiating conversations to check attention and manage break times. The instructor could say, “Hey, are you still with me?” to engage students and share anecdotes related to the lecture, their major, or hobbies (P6, P8). Lecture videos could also pause automatically and ask, “Are you ready to move on?” to check engagement (P7). One participant (P2) suggested embedding inspirational or fun videos upon request, allowing students to set a time limit (e.g., 10 minutes) to help them refocus before returning to the lecture.

%% file: sections/4-system.tex
\section{\system{}}
Based on the formative study results, we developed a proof-of-concept prototype of \system{}. This section presents the system interface and describes the implementation of eight key features.

\subsection{System Overview}
\begin{figure}[h]
    \centering
    \includegraphics[width=\columnwidth]{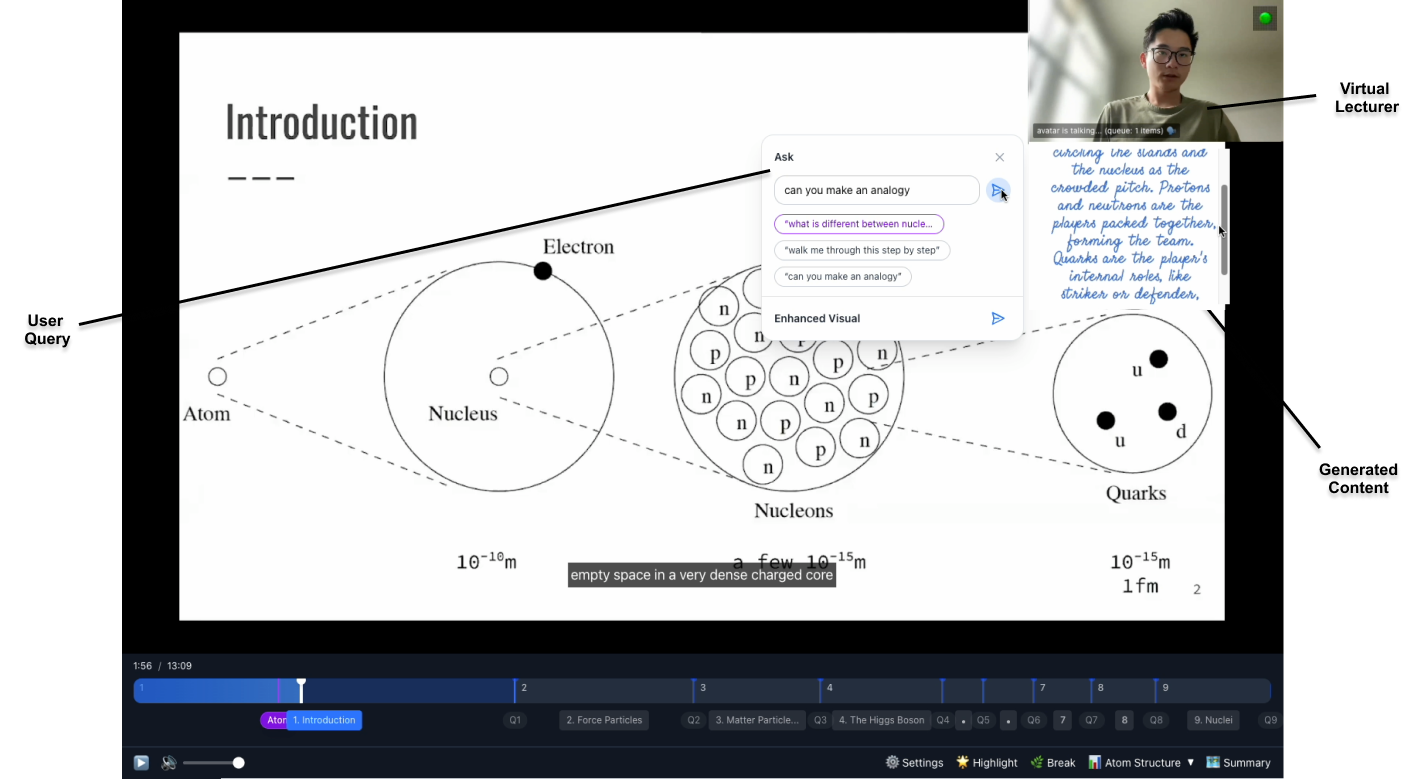}
    \caption{\system{} Interface Overview.}
    \label{fig:system-overview}
    \Description{}
\end{figure}

We developed a web-based interactive video-viewing platform that enhances lecture videos with generative AI to support personalized learning. The platform enables students to ask questions and integrates generated content into the original video through virtual overlays and an AI-clone instructor. As shown in Figure~\ref{fig:system-overview}, the main interface is a familiar video player. At the bottom of the video player, the system annotates the timeline with section titles and quizzes. In the video control bar, in addition to standard features such as play/pause, volume, subtitles, and playback speed, users can click buttons on the bottom right to toggle the automatic highlight feature, initiate a break, and access interactive examples and the summary page.

\subsection{Implementation}
\begin{figure*}[h]
    \centering
    \includegraphics[width=\textwidth]{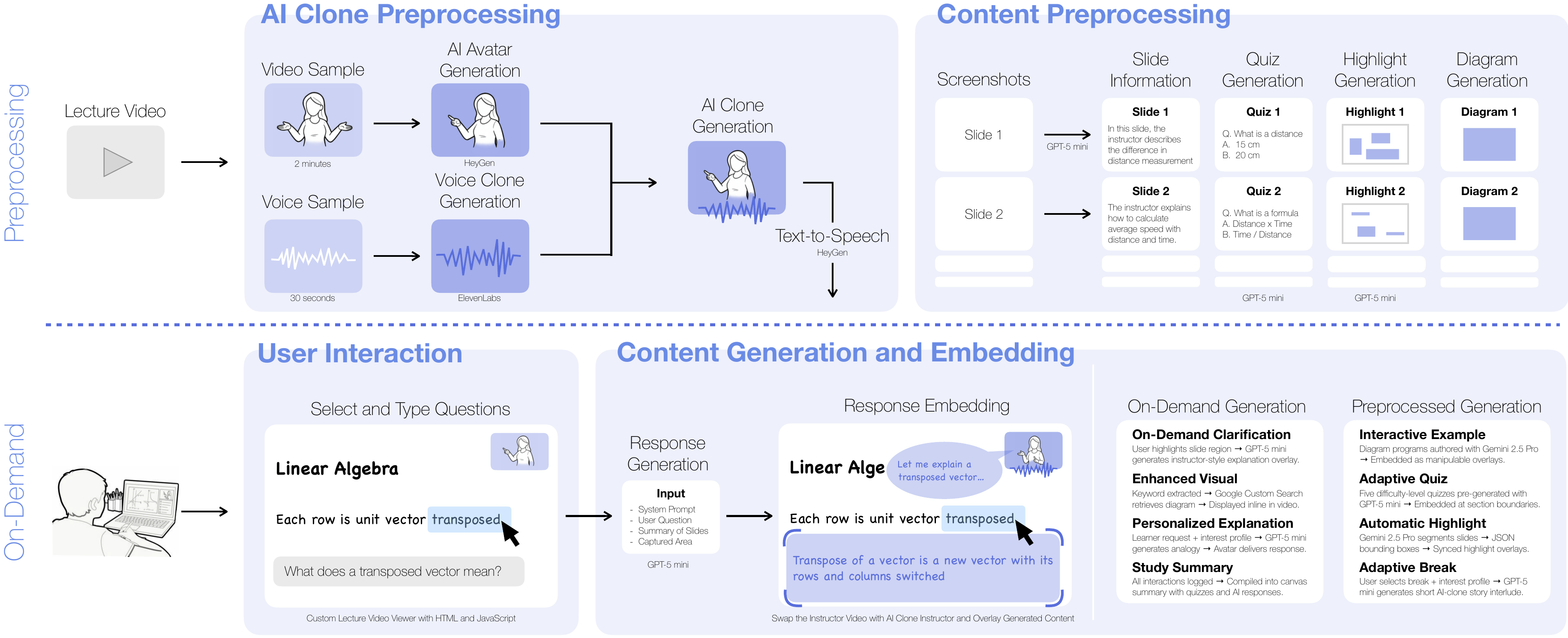}
    \caption{\system{} System Architecture.}
    \label{fig:system-architecture}
    \Description{}
\end{figure*}

The system architecture consists of three main layers: preprocessing, on-demand generation, and content embedding. Figure~\ref{fig:system-architecture} illustrates the overall architecture. 

\subsubsection*{\textbf{Stage 1: Preprocessing}}
In the preprocessing stage, we prepare the instructor representation and lecture content. 

To create an AI-cloned instructor that delivers responses in a consistent manner, we recorded a 2-minute video sample and a 30-second audio clip, which were used for avatar synthesis on HeyGen and voice cloning on ElevenLabs.

For lecture content, we segmented the video into sections using key-frame extraction (via ffmpeg and GPT-5 nano) and aligned transcripts with corresponding slides. Each section was further processed into structured JSON files, which contain textual descriptions, equations, diagrams, and code snippets. At this stage, we also pre-generated supporting materials: adaptive quizzes across five difficulty levels, highlight annotations derived from slide segmentation (Gemini 2.5 Pro), and interactive examples authored via Gemini 2.5 Pro and validated by domain experts for accuracy. Together, these assets form a library of reusable building blocks that can be integrated later.

\subsubsection*{\textbf{Stage 2: On-Demand Generation}}
The on-demand generation stage handles live learner interactions. When students highlight a region or type a question, the system combines contextual inputs---the lecture summary, slide content, user prompt, and optionally the captured area---and queries GPT-5 mini for a tailored response. While waiting, the AI-clone avatar session is prepared, and once the answer is ready, HeyGen text-to-speech animates the instructor to speak, with synchronized text appearing on the empty regions of the slide via Vara.js~\footnote{\href{https://github.com/akzhy/Vara}{https://github.com/akzhy/Vara}}.

This pipeline also supports specific modes of interaction, including enhanced visuals (retrieving relevant images via GPT-5 mini and Google Custom Search), personalized explanations (generating analogies grounded in user-defined student interests), and adaptive breaks (producing short, story-like anecdotes related to the lecture and student interests). These mechanisms ensure that on-demand responses extend beyond factual Q\&A, adapting to both cognitive and affective needs of learners.

\subsubsection*{\textbf{Stage 3: Content Embedding}}
Finally, the embedding stage integrates all preprocessed and generated elements directly into the video interface. Preprocessed content---quizzes, highlights, and interactive examples---is bound to specific timestamps and appears as the lecture progresses. On-demand outputs---clarifications, enhanced visuals, analogies, or generated stories---are inserted dynamically as overlays with both textual and avatar-based auditory delivery.

The summary feature accumulates these interactions: all prompts, responses, highlights, and notes are stored in structured logs and displayed in a post-lecture summary view. This summary is visualized on an infinite canvas, allowing learners to navigate past interactions, replay AI-clone explanations, and revisit adaptive quizzes. By embedding both pre-generated and live content, the system transforms a static lecture video into an interactive, adaptive learning environment.

\subsection{Design and Implementation of Supported Features}
Based on the formative interviews and design elicitation study, we implemented eight interactive features aligned with four learning goals---\textit{improve understanding}, \textit{enhance engagement}, \textit{support review}, and \textit{maintain focus} (Figure~\ref{fig:design-space}). Each feature is tightly integrated with the lecture video through overlays, time-aligned events, and an AI-clone instructor that speaks and writes directly to the learner. Below, we summarize the interactions and implementation details for each feature. We also include detailed LLM prompts in the Appendix to support replication.

\subsubsection*{\textbf{Feature 1: On-Demand Clarification}}
\begin{figure}[h]
\centering
\includegraphics[width=\columnwidth]{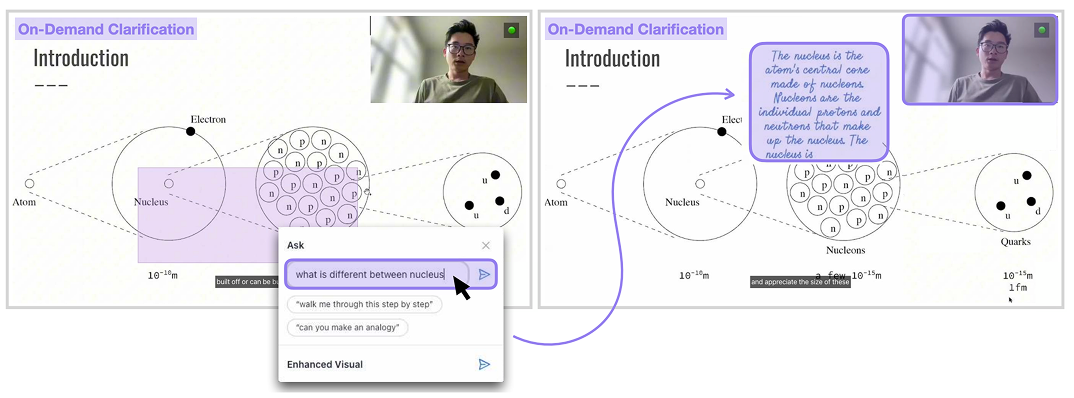}
\caption{On-demand Clarification}
\label{fig:on-demand-clarification}
\Description{}
\end{figure}

On-demand clarification allows users to seek explanations without leaving the video platform. When a region of interest is selected, the video pauses, and a dialogue window appears. For example, as shown in Figure~\ref{fig:on-demand-clarification}, a student confused about “nucleus” versus “nucleons” highlights the slide and asks, \textit{``What is the difference between nucleus and nucleons?''}. The system then generates a concise answer, which is overlaid on the lecture and spoken by the AI-clone instructor with a handwriting effect. By default, the prompt \textit{``Please explain this.''} is pre-filled, but users can also type custom questions. Once the response ends, the text disappears, and the lecture automatically resumes. 

In terms of implementation, when a learner selects a slide region and submits a question, the system packages the system prompt, lecture summary, slide summary, selected region, and user query, and sends them to GPT-5 mini. While the response is being generated, an AI-cloned instructor session is triggered. Once the answer is returned, HeyGen text-to-speech animates the avatar to deliver the explanation, and Vara.js overlays the textual response directly onto the slide.

To determine appropriate overlay locations, we perform grid-based spatial analysis on the slide image. OpenCV extracts content bounding boxes, which are rasterized into a low-resolution occupancy grid. Contiguous unoccupied cells are selected as overlay regions based on adjacency and size. If the generated response exceeds the capacity of the selected region, the system displays the content within that region using a scrollable overlay.

\subsubsection*{\textbf{Feature 2: Enhanced Visual}}
\begin{figure}[h]
\centering
\includegraphics[width=\columnwidth]{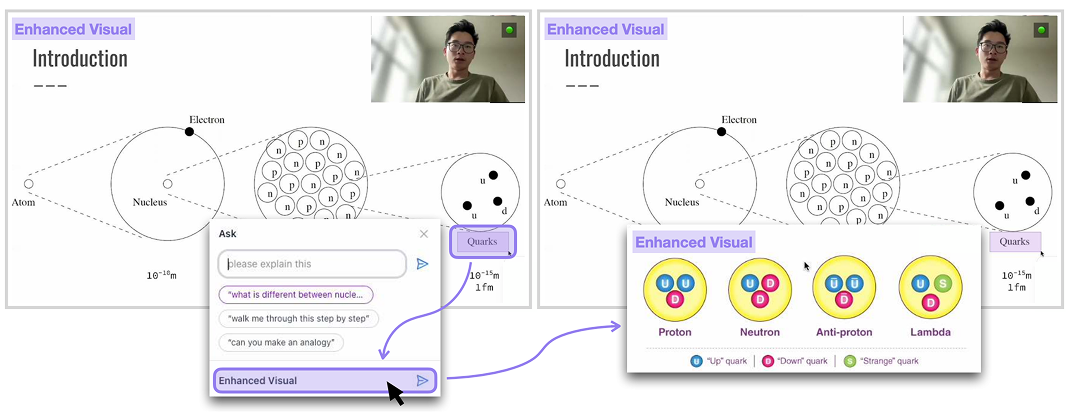}
\caption{Enhanced Visual}
\label{fig:enhanced-visual}
\Description{}
\end{figure}

The enhanced visual feature provides relevant figures through online image search. By pressing the send button in the “Enhanced Visual” block, users can request supporting visuals for a selected keyword. For example, Figure~\ref{fig:enhanced-visual} shows a request for \textit{``Quarks''}, where the system overlays an explanatory diagram of subatomic particles (e.g., proton, neutron, anti-proton, lambda), enriching the lecture with external content.

In the implementation, the system extracts keywords from the selected area using GPT-5 mini and queries Google Custom Search to retrieve relevant images. The retrieved images are displayed in a pop-up visual overlay to augment the original lecture content. The overlay remains visible on the video player until the user manually dismisses it.

\subsubsection*{\textbf{Feature 3: Interactive Example}}
\begin{figure}[h]
\centering
\includegraphics[width=\columnwidth]{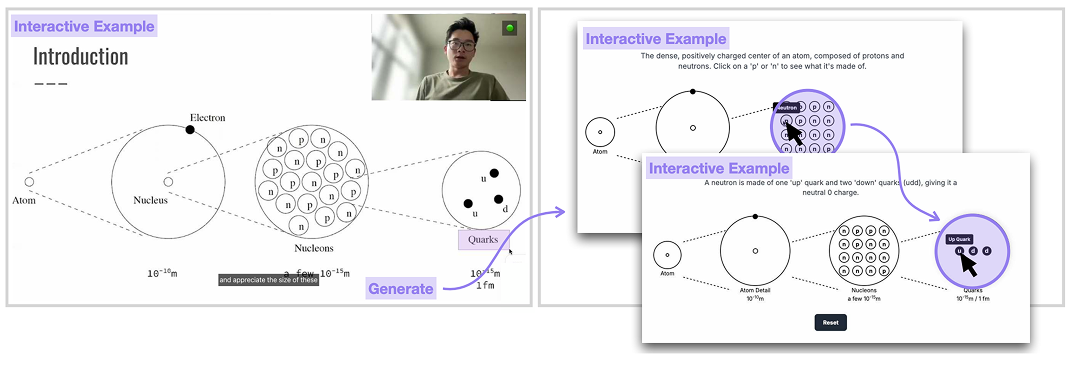}
\caption{Interactive Example}
\label{fig:interactive-example}
\Description{}
\end{figure}

The interactive example feature transforms static diagrams into manipulable elements, making lectures more engaging. As shown in Figure~\ref{fig:interactive-example}, clicking on a neutron reveals its quark composition, allowing students to explore content dynamically. The system can also tailor examples to user information for added relevance. At relevant timestamps, interactive examples automatically appear, giving students a chance to explore concepts right after listening to the instructor's explanation.

For implementation, because generative diagrams can be unreliable when generated on the fly, interactive examples were authored in advance using Gemini 2.5 Pro in canvas mode and verified by domain experts to ensure accuracy. Each example is saved as an HTML file linked to the video source and timestamp, and is embedded into the video timeline as an overlay.

\subsubsection*{\textbf{Feature 4: Personalized Explanation}}
The personalized explanation feature allows learners to request either step-by-step guidance or analogy-based explanations (e.g., \textit{“explain using aerospace examples”}). The dialogue window provides two predefined options: \textit{“Walk me through this step by step”} for detailed breakdowns, and \textit{“Can you make an analogy?”} for explanations grounded in personal interests. When analogies are used, the system connects lecture content to learners’ interests, which are specified in advance and also leveraged by other features such as adaptive breaks. In future iterations, these interests could be inferred automatically from learning histories or everyday activities.

From an implementation perspective, when a learner’s question includes keywords such as “analogy,” the system augments the GPT-5 mini prompt by injecting the learner’s registered interests as contextual constraints. The model then generates a tailored analogy (e.g., comparing atomic structure to a football game), which is delivered by the AI-cloned instructor using the same presentation mechanism as on-demand clarification responses.

\subsubsection*{\textbf{Feature 5: Adaptive Quiz}}
The adaptive quiz appears at the end of each slide to check the student's understanding. A difficulty slider (1: Very Easy --- 5: Very Hard) lets learners adjust the level to their background and confidence. After answering, the system provides immediate feedback and explanations. For example, at medium difficulty, students may recall formulas like $w \cdot x + b$, while at higher levels, they explain the role of the bias term in shifting a perceptron’s decision boundary. In this way, the quiz adapts from basic recall to deeper reasoning, supporting learners across different levels of expertise.

\begin{figure}[h]
\centering
\includegraphics[width=\columnwidth]{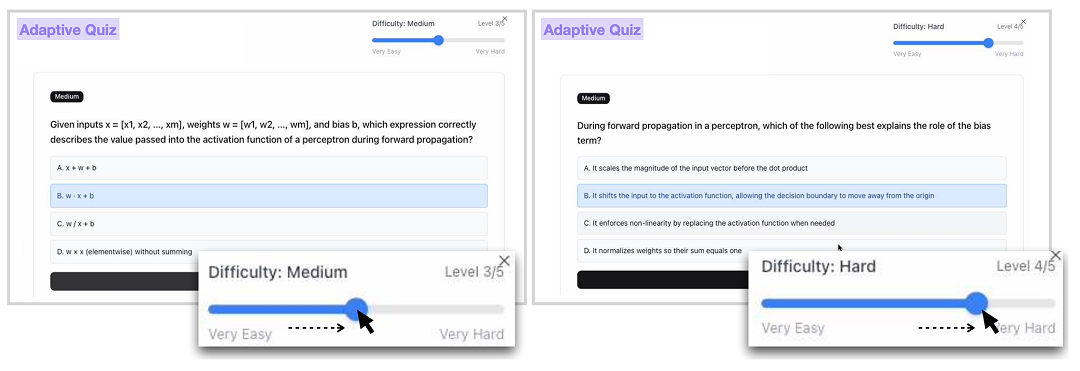}
\caption{Adaptive Quiz}
\label{fig:adaptive-quiz}
\Description{}
\end{figure}

For implementation, each section is preprocessed with adaptive quiz questions spanning five difficulty levels. Questions, options, correct answers, and explanations are generated using GPT-5 mini and stored in JSON format. Learners can adjust a difficulty slider (1–5), dynamically shifting from basic recall (e.g., formula recognition) to higher-level conceptual reasoning (e.g., explaining the role of a bias term).

\subsubsection*{\textbf{Feature 6: Study Summary}}
\begin{figure}[h]
\centering
\includegraphics[width=\columnwidth]{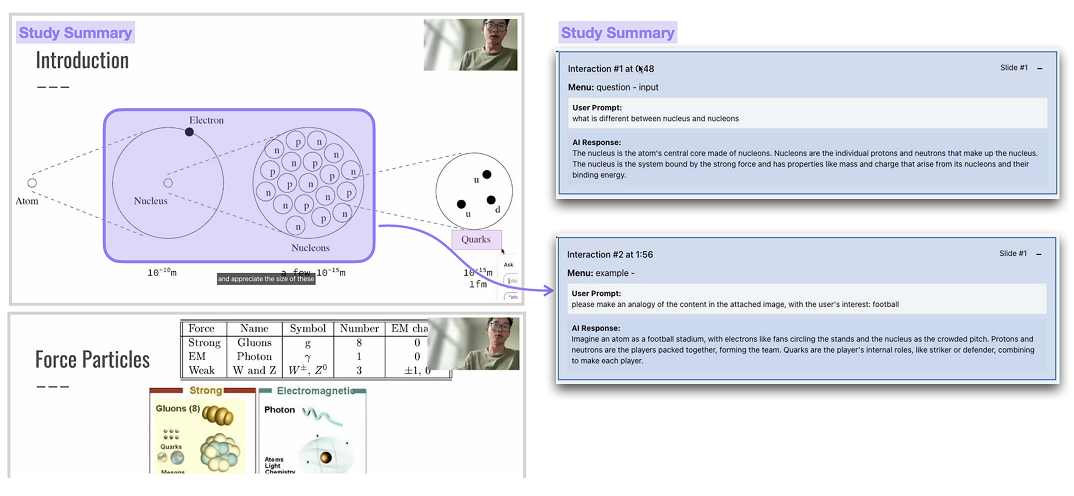}
\caption{Study Summary}
\label{fig:study-summary}
\Description{}
\end{figure}

At the end of a lecture, learners can click the \textit{Summary} button to review their interaction history. This opens a page that integrates section videos with highlights, user prompts, and AI responses. For example, Figure~\ref{fig:study-summary} shows a physics lecture where the system records a question---\textit{“What is different between nucleus and nucleons?”}---and explains that nucleons are protons and neutrons inside the nucleus. The figure also illustrates a personalized analogy request comparing an atom to a football stadium. The study summary consolidates quizzes, notes, prompts, and explanations into a single canvas, helping learners trace their questions, revisit AI-generated explanations, and replay responses without having to rewatch the lecture videos. 

The system logs all user interactions---including questions, selected areas, timestamps, and AI responses---into structured JSON files linked to specific slides. At the end of a lecture, these logs are compiled into a summary page where learners can review screenshots, highlights, quizzes, and their own notes. An infinite canvas visualization serves as a personalized summary map, organizing all interactions for quick review at a later time.

\subsubsection*{\textbf{Feature 7: Automatic Highlight}}
\begin{figure}[h]
\centering
\includegraphics[width=\columnwidth]{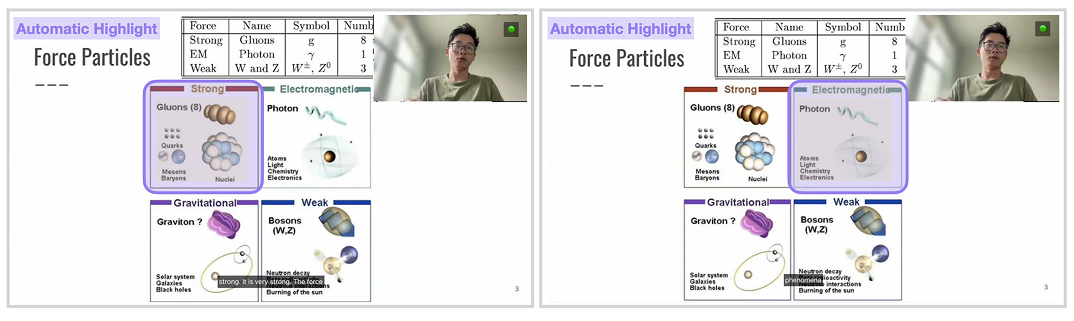}
\caption{Automatic Highlight}
\label{fig:automatic-highlight}
\Description{}
\end{figure}

The automatic highlight feature synchronizes spoken explanations with visual emphasis on lecture slides. As the instructor references a concept, the corresponding region is highlighted, helping learners connect narration to visuals and focus attention. For example, Figure~\ref{fig:automatic-highlight} shows a physics lecture where the system highlights \textit{Gluons} when explaining the strong force and then shifts to \textit{Photon} for the electromagnetic force. By aligning narration with evolving highlights, the system reinforces comprehension and reduces cognitive load. Learners can toggle this feature on or off using the \textit{Highlight} button.

For each key slide, Gemini 2.5 Pro segments the visual layout into blocks and associates them with transcript text and timestamps. This produces a JSON file specifying bounding boxes and time ranges. During playback, relevant regions (e.g., “Gluons” or “Photon” in a physics lecture) are automatically highlighted when the instructor refers to them, aligning narration with visual emphasis.

\subsubsection*{\textbf{Feature 8: Adaptive Break}}
The adaptive break feature allows learners to pause the lecture by clicking the \textit{Break} button and selecting a duration of 1, 3, or 5 minutes. During these pauses, learners are presented with brief, light-weight content that blends lecture topics with their personal interests, supporting motivation and helping them smoothly transition back into the lecture. The virtual lecturer may also use breaks to briefly check comprehension and offer simplified explanations when appropriate.

When a break is requested, the system constructs a GPT-5 mini prompt using the lecture summary, the learner’s registered interest profile, and the selected break duration. The model generates a short narrative, which is delivered by the AI-cloned instructor as temporary interlude content before the lecture resumes.

%% file: sections/5-evaluation.tex

\section{User Study}
To understand the system’s usability and its potential to support student learning and teaching workflows, we conducted two studies: 1) a user study with students and 2) an expert review with instructors. In this section, we first describe the user study. We conducted a within-subjects comparative study with twelve students (P1-P12) from our institute to assess the system’s usability and explore how its interaction features support filling knowledge gaps, enhancing engagement, checking their understanding, and staying focused. 

\subsection{Participants}
We recruited twelve participants (7 females, 5 males), aged 22–28 ($M = 24.67$, $SD = 1.97$), through a flyer posted on a bulletin board and a student email list at our institution. Participants reported an average of 7.9 hours of lecture video use per week ($SD = 11.3$), and their backgrounds included computer science (3), human-computer interaction (5), electrical engineering (2), chemistry (1), and mathematics (1). They all had previous experience with interactive lecture video platforms, such as Canvas, Coursera, and Khan academy, and have watched lecture videos across a wide range of subjects, most often in technical domains such as computer science (e.g., machine learning, systems, networks, graphics, programming), engineering (e.g., electrical, electronics, quantum physics), and mathematics. Several participants mentioned design-related interests (e.g., UX, art, game design, and 3D modeling), while others explored interests outside their primary field of study, such as music production, psychology, or sociology.

\subsection{Method}
To examine the influence of specific features on the learning experience, we compared our system with a baseline interactive lecture video watching platform. 
\begin{itemize}
    \item \textbf{Baseline}: video player containing pre-defined quizzes, similar to existing interactive lecture video platforms.  
    \item \textbf{\system{}}: An LLM-powered interactive lecture video platform with AI clone instructors.
\end{itemize}

\subsubsection{Materials}
For content, we prepared two lecture videos at the introductory level: (1) \textit{MIT Introduction to Deep Learning - Perceptron} by Prof. Alexander Amini~\footnote{\url{https://introtodeeplearning.com}}, and (2) \textit{MIT Introduction to Nuclear and Particle Physics - Particle} by Prof. Markus Klute~\footnote{\url{https://ocw.mit.edu/courses/8-701-introduction-to-nuclear-and-particle-physics-fall-2020/resources/lecture-0-5/}}. To generate the AI-cloned instructor without appropriating the original lecturer’s likeness or intellectual property, we re-recorded the lecture videos using their slides and scripts as they were, solely for research purposes. We also manually corrected minor errors in the automatic highlights and interactive examples, so that participants could focus on the interaction aspects rather than being distracted by system errors.

\subsubsection{Procedure}
Upon arrival, participants were briefed about the study and completed a demographic survey on their prior learning experiences and challenges with lecture videos. We then provided a walkthrough of the system, introducing and demonstrating each feature. Each participant watched one video in the baseline condition and the other in our system, with video-condition assignments fully counterbalanced. With participants’ consent, we logged all on-screen interactions and audio during the study. To evaluate learning gain, participants also completed a pre-test and post-test on the lecture content for each video. Afterward, they filled out questionnaires evaluating their overall learning experience with each system, followed by a brief semi-structured interview to elaborate on their responses. After both sessions, we conducted a post-hoc interview to gather qualitative insights. The study lasted 90 minutes in total, and participants were compensated at a rate of 15 USD per hour, resulting in a payment of 22.50 USD each.

\subsubsection{Measurement}
We collected participants' perceived task load using NASA-TLX~\cite{hart1986nasa} and system usability using the System Usability Scale (SUS)~\cite{brooke1996sus} questionnaires. In addition, we gathered participants' ratings on our custom questionnaires measuring learning satisfaction, understanding, engagement, review, focus, and adoption (see Appendix Section~\ref{section:custom-questionnaire} Table~\ref{tab:custom-questionnaire} for details). In the \system{} condition, we also asked participants to rate the perceived usefulness of each feature. We analyzed learning outcomes by comparing pre- and post-test scores between the two conditions. Lastly, we conducted a thematic analysis of post-hoc interview responses to extract qualitative insights.

\subsection{Result}

\subsubsection{Task Load and Usability Ratings}
\begin{figure}[h]
    \centering
    \includegraphics[width=\columnwidth]{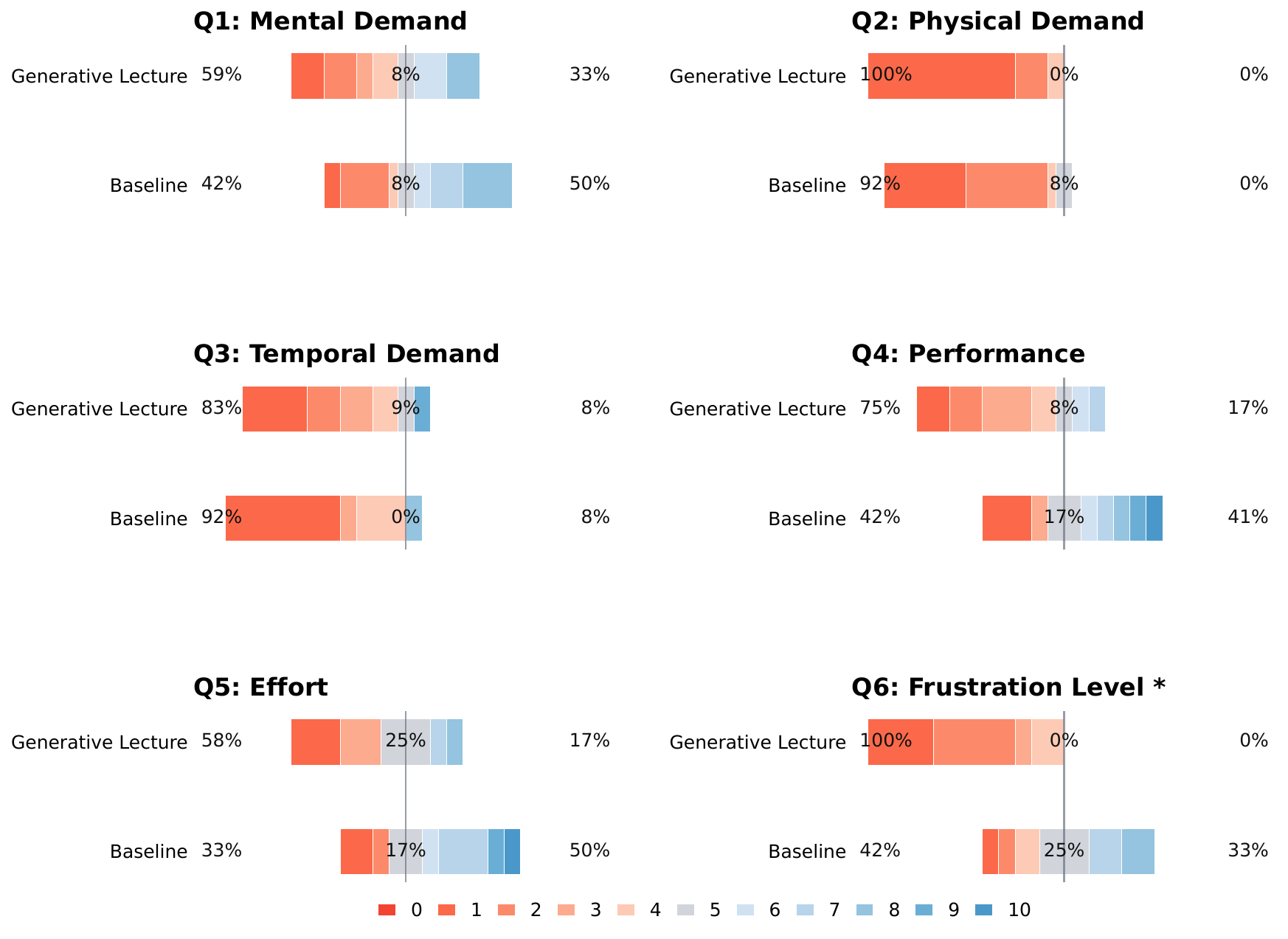}
    \caption{Task Load Questionnaire Result (NASA-TLX). The performance scale is opposite.}
    \label{fig:study_nasa_tlx}
    \Description{}
\end{figure}

\begin{figure}[h]
    \centering
    \includegraphics[width=\columnwidth]{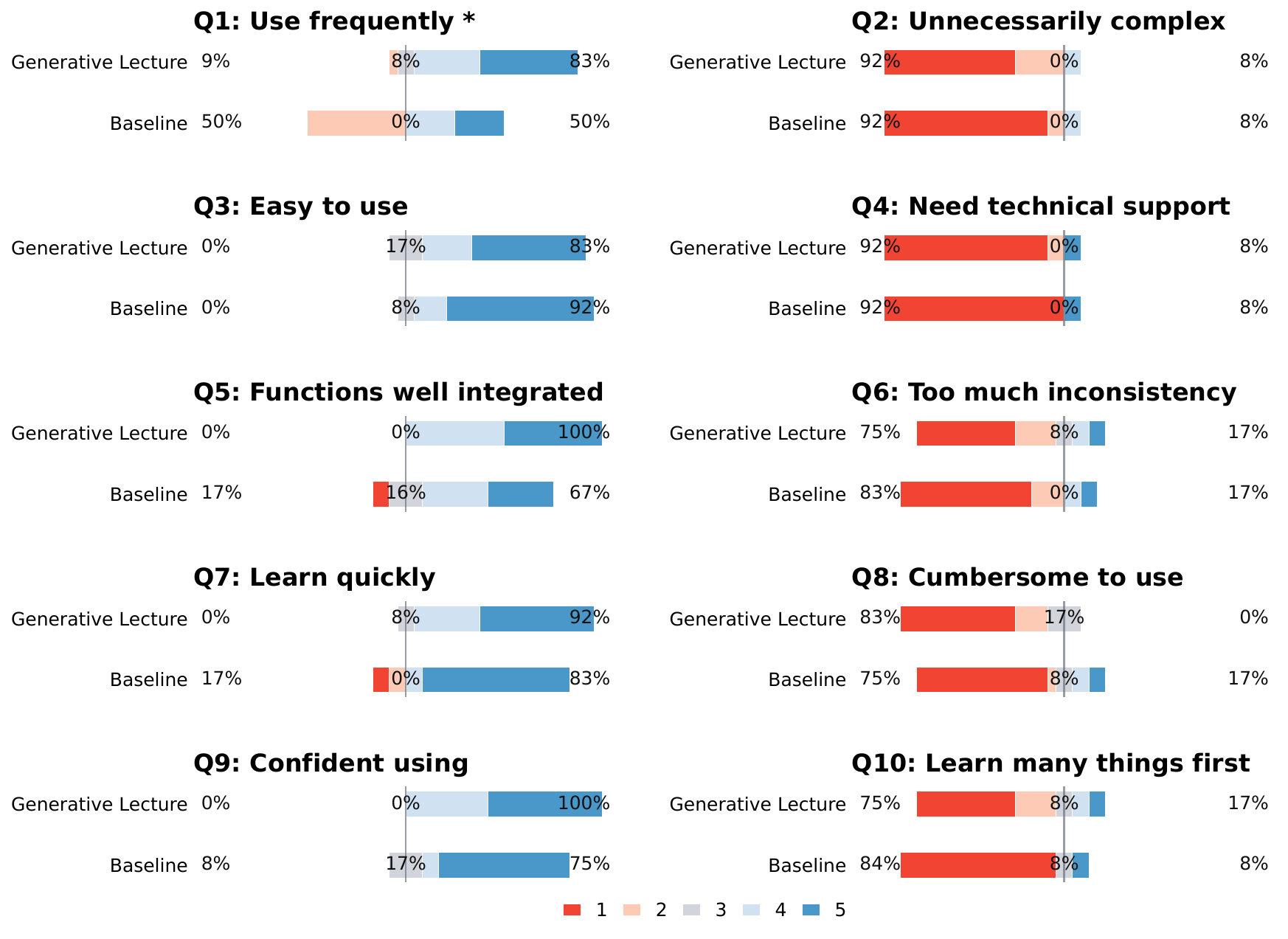}
    \caption{System Usability Questionnaire Result (SUS)}
    \label{fig:study_sus}
    \Description{}
\end{figure}

Figure~\ref{fig:study_nasa_tlx} and Figure~\ref{fig:study_sus} compare participants’ perceived task load and system usability across the two systems. In the NASA-TLX results, frustration was significantly lower with \system{} ($M = 2.08$, $SD = 1.08$) than with the baseline ($M = 5.00$, $SD = 2.22$), with $t(11) = -4.52$, $p < .001^{***}$. While the other dimensions did not differ, mental demand and effort tended to be higher in the baseline condition. These findings suggest that participants experienced less strain and were able to study more smoothly with our system.

Turning to the System Usability Scale (Figure~\ref{fig:study_sus} and Table~\ref{tab:SUSResult}), both systems achieved excellent usability ratings over eighty. However, participants indicated a stronger intention to use \system{} more frequently ($M = 4.25$, $SD = 0.97$) compared to the baseline ($M = 3.25$, $SD = 1.36$), $t(11) = 2.71$, $p = .02^{*}$. Together, these results highlight not only the reduced cognitive strain of \system{} but also its greater appeal for continued use.

Together, these results indicate that \system{} reduced participants’ frustration and cognitive strain, while also being rated as more likely to be used in practice compared to the baseline.

\begin{table}[h]
\centering
\begin{tabular}{lccc}
\toprule
\textbf{Interface} & \textbf{Mean (SD)} & \textbf{Rating} \\
\midrule
\system{} & 84.38(8.99) & Excellent\\
Baseline & 81.46(13.96) & Excellent\\
\bottomrule
\end{tabular}
\caption{SUS Score (p<0.005)}
\label{tab:SUSResult}
\end{table}

\subsubsection{Pre- and Post-Quiz Results}

As for learning gains (Table~\ref{tab:learning-gain}), we found no significant difference between the baseline and \system{} conditions ($t(11) = -0.29$, $p = .78$). On a 0-100 scale, the baseline condition improved by $M = 50.83$ ($SD = 21.51$), while \system{} improved by $M = 48.33$ ($SD = 22.90$). 

We also found no effect of subject domain (machine learning ($t(11) = 1.08$, $p = .31$), particle physics ($t(11) = -0.65$, $p = .53$)) ($t(11) = 1.28$, $p = .23$). 

Overall, there was no evidence of system- or subject-level differences in learning gain. A longer-term study may be needed to capture delayed or cumulative effects.

\begin{table}[t]
\centering
\begin{tabular}{lccc}
\toprule
{Condition} & \textbf{Mean (SD)} \\
\midrule
Baseline           & 50.83 (21.51) \\
Generative Lecture & 48.33 (22.90) \\
\bottomrule
\end{tabular}
\caption{Learning gain (Post--Pre Quiz) by condition.}
\label{tab:learning-gain}
\end{table}

\subsubsection{Custom Questionnaire Ratings}

\begin{figure}[h]
    \centering
    \includegraphics[width=\columnwidth]{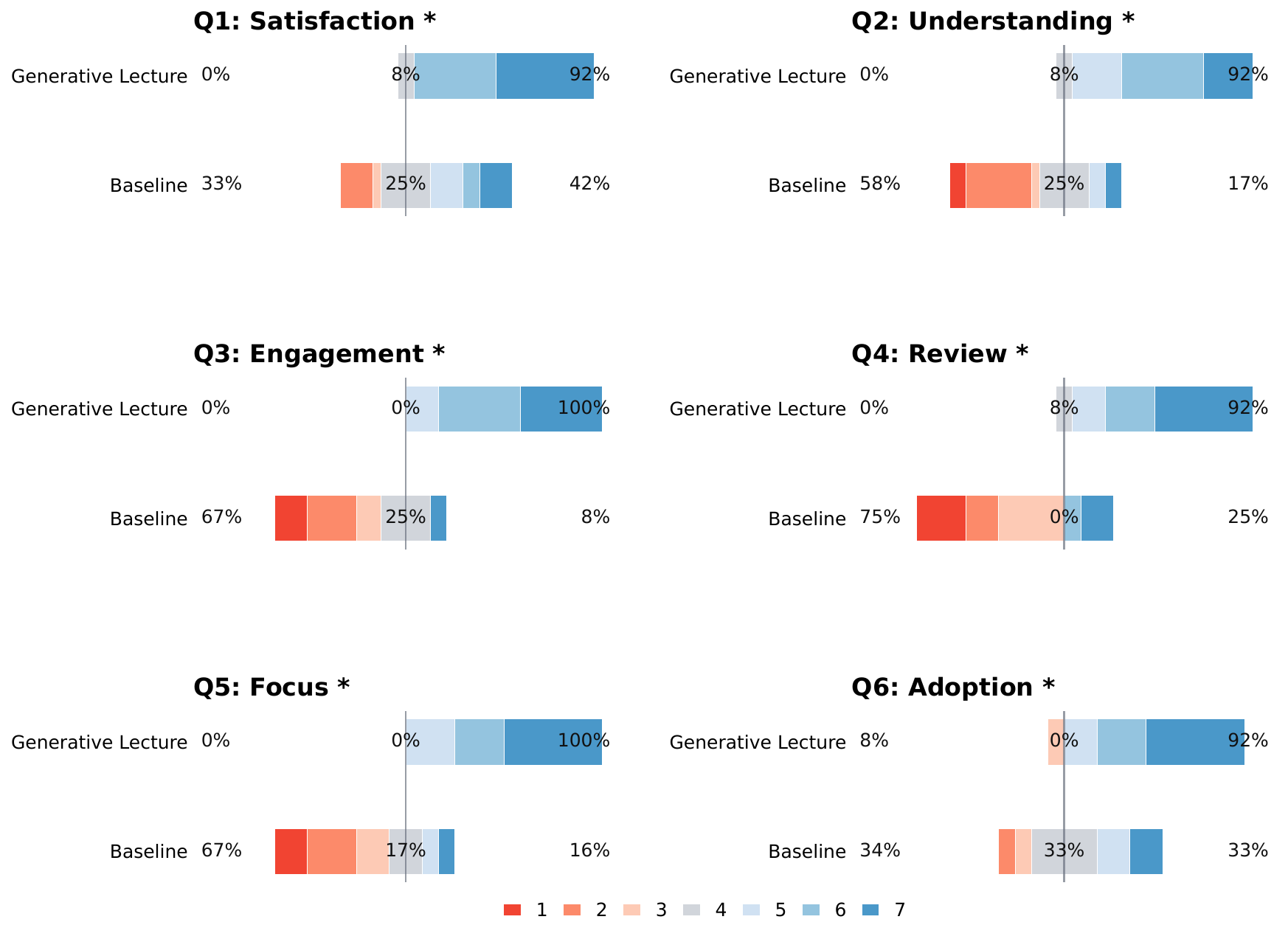}
    \caption{Custom Questionnaire Result}
    \label{fig:study_general_eval}
    \Description{}
\end{figure}

As shown in Figure~\ref{fig:study_general_eval}, participants were overall more satisfied with \system{} ($M = 6.33$, $SD = 0.89$) than with the baseline ($M = 4.33$, $SD = 1.72$), $t(11) = 3.94$, $p = .002^{**}$. They also gave significantly higher ratings across all questions related to understanding, engagement, review, focus, and adoption (Appendix Table~\ref{tab:custom-questionnaire}). The highest ratings were for engagement, with a mean score of $6.25$ ($SD = 0.75$), $t(11) = 7.29$, $p < .001^{***}$. Interestingly, the lowest ratings were for understanding. A possible explanation is that some participants felt the system offered too many features, which at times distracted them from focusing on the lecture content.

\subsection{Feature Evaluation Result}


Figure \ref{fig:study_feature_evaluation} presents participants' evaluations of our eight features. Their comments on each feature are summarized below:

\begin{figure}[h]
    \centering
    \includegraphics[width=\columnwidth]{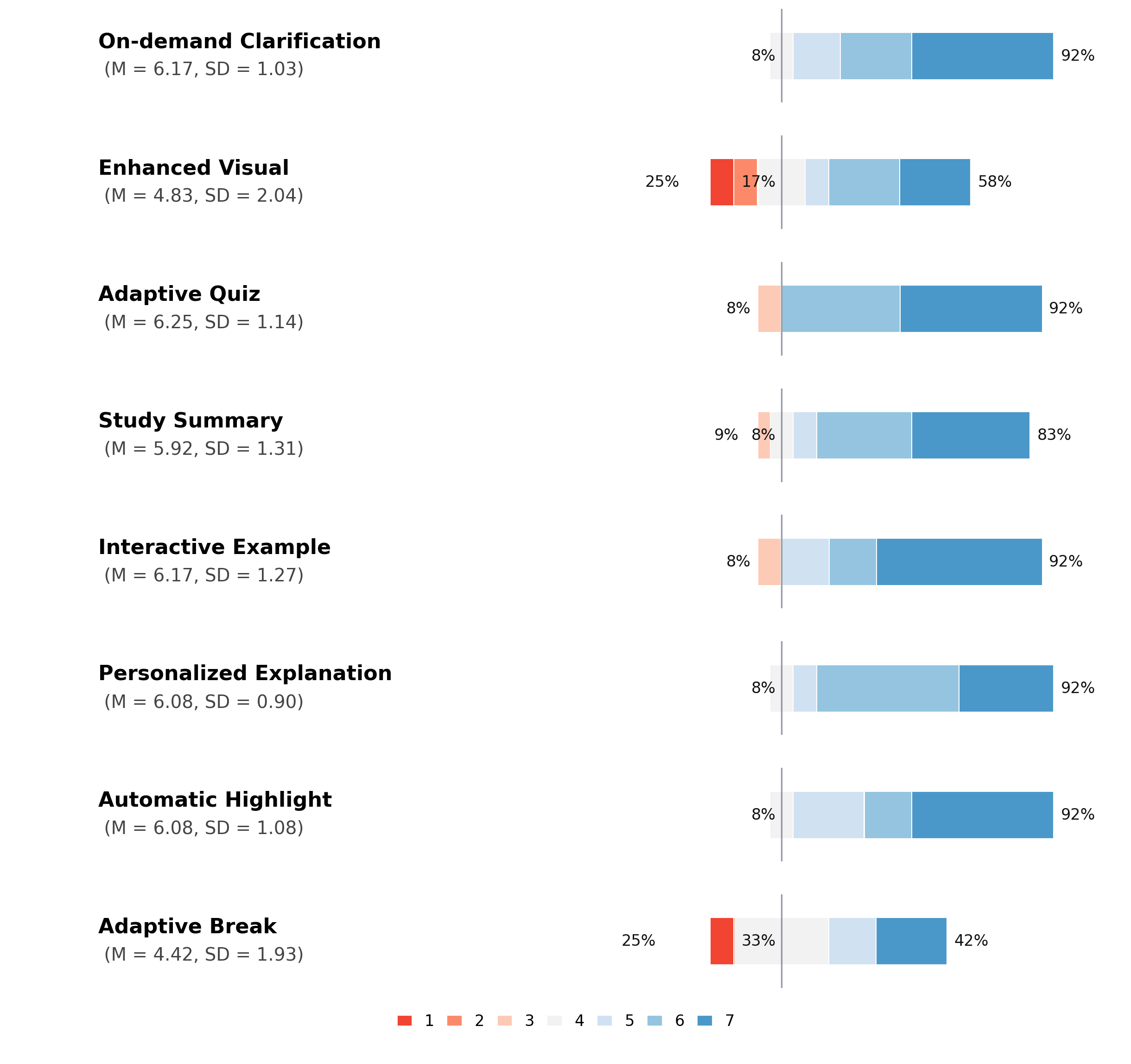}
    \caption{Feature Evaluation Result}
    \label{fig:study_feature_evaluation}
    \Description{}
\end{figure}

\subsubsection{On-Demand Clarification}
Participants P6, P7, P8, and P12 appreciated the convenience of being able to ask questions within the same platform. P8 noted that \textit{``it's easier for me to ask a question in one tab and I don’t have to switch tabs. I can see what is taught, and I don’t need to explain everything to ChatGPT''}. Similarly, P3 shared that \textit{``the idea of I can search for more information makes me focused, it’s a new experience for me''}. However, P10 stated that this feature would not replace their use of ChatGPT, since \textit{``my ChatGPT knows me more, it always explains the question in a way that I can easily understand''}.

\subsubsection{Enhanced Visual}
P8 commented that \textit{``I am a visual learner, so I really want a visual explanation, but I want it to be correct''}. P1 also expressed a similar concern, saying that "the first image returned by the search engine might not be the best answer in terms of the information level. You may want a library for STEM terms".

\subsubsection{Interactive Example}
After viewing all generated examples, P3 stated that they would \textit{``favor exploratory examples over explainable examples''}. Both P3 and P10 suggested that the Higgs boson decay could be used as a more engaging interactive example for the Particle Physics lecture.

\subsubsection{Personalized Explanation}
P8 remarked that \textit{``For people with no clue, it will help them relate''}. However, they recognized that some generated analogies were inaccurate. For example, the analogy comparing machine learning to basketball mentioned that there are 10 people on the court, whereas there are actually 5 players per team.

\subsubsection{Adaptive Quiz}
We observed that participants typically challenged themselves by progressing to higher difficulty levels after correctly answering medium-level questions. Some participants, lacking prior knowledge, chose to remain at the lowest level of difficulty (1/5). P4 mentioned that \textit{``I got lost for a bit, but the explanation of the quiz helped''}.

\subsubsection{Study Summary}
P12 described the study summary feature as \textit{``good to have''}. Given the limited duration of the study, most participants did not provide extensive feedback on this feature.

\subsubsection{Automatic Highlight} 
All participants agreed that the automatic highlight feature supported their focus, particularly when the instructor deviated from the slide order during explanations. None of the participants manually disabled this feature, even after it was pointed out during the walkthrough. P10 shared that \textit{``the highlight was my favorite feature because the mouse move in the original recording is very rare. I want more of this''}. 

\subsubsection{Adaptive Break}
Not all participants tried the break feature, as many felt a 10-minute lecture did not warrant a break. Among those who did try it, P11 mentioned that the break story about particle physics and algebra "doesn't make sense. It's just replacing words with my algebra terms".

\subsection{Learning Outcome}
Participants using the baseline platform showed an average improvement of $5.08$ points ($SD = 2.15$) in the post-test, while those using the \system{} platform improved by $4.83$ points ($SD = 2.29$). A Welch's t-test revealed no statistically significant difference between the groups ($t = -0.28$, $p = 0.79$). 

\subsection{Qualitative Feedback}

\subsubsection{Supporting Follow-up Questions}
P3 stated that \textit{``the system feels like a mix of watching and conversation mode with the auto resume. What if I want to ask a follow-up question''}. They suggested providing manual control to switch back to watching mode or keeping the dialogue window always accessible.

\subsubsection{Personalization beyond Simple Keywords}
Participants appreciated the idea of personalized explanations, even though they were not always accurate. Several suggested incorporating content more deeply tailored to their personal interests, rather than relying on simple keyword-based subjects. P7 remarked that if they asked about applications, \textit{``it would be great to hear about the application of Deep Learning in basketball''}. P8 found the Higgs Boson pizza image in the Particle Physics lecture particularly engaging, interpreting it as personalized to their interest in food. In addition, participants suggested that \textit{``the system can give a personal introduction for each student at the beginning of the lecture, combined with their interest''}.

\subsubsection{Social Presence of AI-Cloned Instructors}
Participants noted that having an AI clone of the instructor felt like having company. P10 emphasized that \textit{``It makes me feel like I’m not falling behind, because the avatar is there to answer my questions.''}. While most participants enjoyed the avatar responding to them, some shared concerns. P3 expressed concern that the AI clone may initially make students feel closer, but because it is not the instructors themselves, it could ultimately increase the distance between students and instructors.

\section{Expert Interview}
We conducted an expert review with five university lecturers (E1-E5) who had prior experience teaching through lecture videos. The review aimed to collect expert feedback on the system's applicability as a learning material for their classes. In addition, we gathered the instructors' perspectives on how they envisioned AI clones interacting with students.

\subsection{Participants}
We recruited participants through university email lists. Participants (3 females, 2 males), aged 29–53 ($M = 39.40$, $SD = 4.67$), had 2–15 years of teaching experience and 2 months to 2 years of experience creating lecture videos ($M = 12.80$, $SD = 11.01$). Recruited participants primarily taught computing and design-related subjects, such as introductory computer science, web and software development (JavaScript, NodeJS, Express, C\#), creative coding, data visualization, and physical computing. They used various video lecture platforms, such as Coursera, YouTube, Zoom, and Canvas.

\subsection{Procedure}
After explaining the purpose of the interview and collecting demographic information, we presented a brief walkthrough of the system’s concept and key features. Participants then engaged with the system using two sample lecture videos from the prior student usability study in a think-aloud manner. We subsequently conducted a semi-structured interview focusing on the system’s applicability as a learning material. 
We asked how they envisioned their students using the system for their lecture videos, how an AI clone of themselves should interact with learners, and what risks they saw in terms of accuracy and student reliance. We also discussed what kinds of data the system should collect to create a meaningful feedback loop for their teaching practice.

For analysis, we conducted an inductive thematic analysis. Two researchers independently reviewed transcripts and identified recurring patterns related to pedagogical benefits, concerns, and expectations for system use. Themes were then collaboratively refined through discussion.

\subsection{Results and Findings}
The expert interviews revealed a range of reactions to the proposed system, reflecting instructors’ expertise and teaching experience. The themes below highlight both the opportunities they saw and the concerns they raised.

\subsubsection{Expandable Lecture Content Beyond Static Lecture Videos}
Instructors acknowledged that while traditional lecture videos are efficient, they often fall short when course content needs to evolve. One participant noted that videos covering \textit{``state-of-the-art or practical content''} may become outdated and require updates (E4). Rather than re-recording entire lectures, some saw potential in using AI to supplement static videos with clarifications or additions, allowing materials to remain relevant over time. 

\subsubsection{Use of AI to Offload Repetitive Clarification Tasks}
Beyond content flexibility, instructors also viewed AI augmentation as a way to reduce their cognitive load by automatically handling frequently asked questions. One instructor explained, \textit{``Given variance (of the student background in introductory class), it’s hard to anticipate students’ questions [...] but having AI handle the common ones would be hugely beneficial''} (E1). Several participants emphasized that providing on-demand clarification could make pre-recorded videos more effective as supplementary materials. 

\subsubsection{Mixed Feelings About AI Clone}
While participants welcomed features such as on-demand explanations and highlights, three out of five instructors expressed discomfort with hyper-realistic AI avatars. One said, \textit{``It looks like the instructor is possessed by the computer [...] It might say something I don’t believe in''} (E1). Another added, \textit{``I don’t want students to 100\% trust the AI answer with my face''} (E3). Nevertheless, four participants agreed that having a visible instructor face made the interaction more engaging. To address this concern, instructors suggested making the AI identity explicit through visual tags or using stylized representations, such as cartoon or robotic avatars (E1, E3), to help students clearly distinguish the AI agent from the human instructor.

\subsubsection{Conditional Trust in AI Explanations} 
Instructors linked their acceptance of AI augmentation to the type of course content. They considered AI suitable for introductory topics: \textit{``Accuracy was okay [...] in something like Intro to Programming, ChatGPT could be fine. In fact, they are really good at that''} (E1). However, for advanced or specialized material, instructors stressed the need for oversight. As one explained, \textit{``For more advanced or practical lecture content, I need to check the quality''} (E4). This suggests that while AI-augmented videos may support foundational learning with minimal intervention, their use in higher-level instruction would require more careful policy and instructor review to ensure accuracy and trustworthiness.

\subsubsection{Preference for Summary Insights Over Monitoring} 
A persistent challenge with lecture videos is the lack of real-time feedback on what students understand. In the demographic survey, several instructors mentioned that asynchronous formats make it difficult to gauge comprehension or identify where students struggle. In the interviews, instructors saw potential in AI to surface patterns of confusion rather than track individual interactions. One requested \textit{``a high-level summary, top 10 questions''} (E1), while another said, \textit{``Knowing which keywords they use a lot might show me where I missed it.''} (E4). They viewed such aggregated insights as more valuable for instructors than detailed logs. Rather than monitoring individual interactions, they saw AI augmentation as a way to support reflective teaching and flipped learning.

%% file: sections/6-discussion.tex
\section{Discussion \& Future Work}

While our study suggests the promise of bi-directional lecture videos, several limitations and opportunities for future work remain.

\subsection{Multifaceted Personalization in Learning}
\system{} illustrates how personalization can enrich lecture videos by tailoring explanations to individual learners. In our implementation, this took the form of analogy-based explanations that built on students’ existing backgrounds or interests, represented as short keywords. Such personalization can lower barriers to understanding and mirror how human instructors naturally adapt their examples. However, the current feature does not reflect the multifaceted characteristics of personalization. Future work can extend this idea to additional dimensions, such as adjusting the teaching style (e.g., theoretical derivations vs. applied, real-world explanations), modifying the order and pacing of topics, or offering more conversational discussions for learning through questions.

\subsection{Limited and Homogeneous Participant Samples}
Our evaluations involved a relatively small number of students and instructors, primarily from STEM backgrounds and with prior experience using lecture videos. This homogeneity limits the generalizability of our findings to broader learning contexts. Expanding to more diverse groups---including humanities learners or instructors with different teaching philosophies---will help uncover how disciplinary or cultural differences shape the use and perception of AI-augmented lecture videos. Therefore, future studies should recruit more varied populations to strengthen external validity.

\subsection{Short-Term Study and Learning Outcomes}
We measured pre- and post-test performance in a single-session study and found no significant differences in learning gains between the baseline and our system. While our study results still provided useful insight in terms of interactions, they may not capture the delayed or cumulative benefits of interactive lectures. Longitudinal studies are necessary to investigate retention, transfer, and metacognitive outcomes over several weeks or months.

\subsection{Trust, Accuracy, and Reliability of AI Content}
Ensuring reliability is critical for integrating AI into high-stakes educational contexts and for building long-term trust among both learners and teachers. Participants and instructors raised concerns about the accuracy of AI-generated visuals and explanations. They noted that even minor mistakes could pull learners out of the flow and make them question whether they could rely on the system. The reliability of third-party services, such as image search, further complicates this issue. Future work should explore validation mechanisms, transparency features such as citations or confidence indicators, and workflows that enable instructors to curate or approve generated content. 

\subsection{Learners’ Relationship with AI Clones}
AI-cloned instructors helped create a sense of continuity, making it feel as though learners were still in conversation with the lecturer rather than switching to external tools. This made interactions smoother and less disruptive. At the same time, participants questioned the credibility and sincerity of clones, saying that small mistakes felt more serious when delivered by an instructor-like figure. Future work should investigate how learners develop trust in AI clones over time and how clones can best complement, rather than replace, human teachers.

%% file: sections/7-conclusion.tex
\section{Conclusion}

In this paper, we presented \system{}, a concept that makes lecture videos interactive by embedding AI-clone instructors and generative content. Through formative studies, we identified eight key features that support students’ understanding, engagement, review, and focus. User evaluations show that our system reduces frustration, enhances engagement, and offers meaningful personalization, though immediate learning gains were not significantly different from baseline. Instructors and students also raised concerns about trust and accuracy, highlighting the need for careful design and transparency. Overall, this work demonstrates the potential of generative AI to turn static videos into adaptive, conversational learning partners, informing the design of future AI-powered educational tools.

%% file: sections/appendix.tex
\appendix

\section{LLM Prompts}

\subsection{Video Segmentation}~\label{video-segmentation}
\begin{lstlisting}[label={lst:particles_json}]
You are an expert at comparing lecture slides to determine if they represent the same underlying content.

IMPORTANT: Two frames show the SAME SLIDE if they have the same core content, even if:
- Annotations have been added (hand-drawn marks, highlighting, underlining)
- A human appears in frame (instructor video overlay or physical presence)
- The cursor has moved
- Minor animations or transitions are occurring
- Text is being progressively revealed (but the slide topic is the same)

Two frames show DIFFERENT SLIDES if:
- The main topic or title has changed
- The slide layout is fundamentally different
- It's a completely new set of informa(Appendix~\ref{quiz-generation})tion
- Major content sections have changed (not just annotations)

Return a JSON object:
{
  "isSameSlide": true/false,
  "confidence": 0.0-1.0,
  "reason": "Explanation of your decision",
  "contentChange": {
    "type": "annotation|human_motion|cursor|new_slide|transition",
    "description": "What changed between the frames"
  }
}
\end{lstlisting}

\subsection{Slide Information Extraction}~\label{slide-information-extraction}
\begin{lstlisting}[label={lst:particles_json}]
You are an expert at analyzing lecture slides and educational content.
            
Your task is to identify the CORE CONTENT of the slide, ignoring temporary overlays like:
- Hand annotations, drawings, or highlighting
- Cursor movements or pointer indicators
- Human presence (instructor's video overlay or physical presence)
- Temporary pop-ups or notifications

Focus on the underlying slide structure and permanent content.

Return a JSON object with this structure:
{
  "title": "Main slide title or topic",
  "mainTopics": ["key topic 1", "key topic 2", ...],
  "hasHumanPresence": true/false,
  "hasAnnotations": true/false,
  "contentFingerprint": "A unique identifier based on the core slide content (ignore annotations)",
  "description": "Brief description of the slide's main content"
}
\end{lstlisting}

\subsection{Quiz Generation}~\label{quiz-generation}
\begin{lstlisting}[label={lst:particles_json}]

You are an educational quiz generator. Create high-quality, pedagogically sound quiz questions based on educational content. 

Your questions should:
- Test understanding, not just memorization
- Be clear and unambiguous
- Have explanations that help reinforce learning
- Cover the most important concepts from the content
- Be appropriate for the specified difficulty level (1=very easy, 5=very hard)

Based on the following educational content, generate exactly ${questionsPerSection} quiz questions.

**SLIDE CONTENT:**
Title: ${slideData.title}
Main Concepts: ${slideData.content.mainConcepts.join(', ')}
Key Points: ${slideData.content.keyPoints.join(', ')}
${slideData.content.equations ? `Equations: ${slideData.content.equations.join(', ')}` : ''}
${slideData.content.diagrams ? `Diagrams: ${slideData.content.diagrams.join(', ')}` : ''}

**TRANSCRIPT:**
${slideData.transcript}

**REQUIREMENTS:**
- Generate exactly ${questionsPerSection} questions
- Difficulty level: ${typeof difficulty === 'number' ? `${difficulty}/5 (${difficulty === 1 ? 'very easy - basic recall' : difficulty === 2 ? 'easy - simple understanding' : difficulty === 3 ? 'medium - application' : difficulty === 4 ? 'hard - analysis' : 'very hard - synthesis/evaluation'})` : difficulty}
- Question types to use: ${questionTypes.join(', ')}
- Mix question types if multiple types specified
- Focus on the most important concepts
- Ensure questions test understanding, not just memorization

**RESPONSE FORMAT:**
Return a JSON object with this exact structure:
{
  "questions": [
    {
      "type": "multiple-choice",
      "question": "Question text here?",
      "options": ["Option A", "Option B", "Option C", "Option D"],
      "correctAnswer": "Option A",
      "explanation": "Detailed explanation of why this is correct and how it relates to the slide content.",
      "difficulty": "medium"
    },
    {
      "type": "true-false",
      "question": "Statement to evaluate as true or false.",
      "options": [],
      "correctAnswer": "True",
      "explanation": "Explanation of the concept and why the statement is true/false.",
      "difficulty": "easy"
    },
    {
      "type": "fill-blank",
      "question": "Complete this statement: Linear transformations preserve _____ and scalar multiplication.",
      "options": [],
      "correctAnswer": "addition",
      "explanation": "Linear transformations must preserve both addition and scalar multiplication by definition.",
      "difficulty": "medium"
    }
  ]
}

**QUESTION TYPE GUIDELINES:**
- multiple-choice: 4 options, only one correct (provide options array)
- true-false: Statement that can be evaluated as true or false (use empty options array)
- fill-blank: Complete a sentence or equation, answer should be a single word or short phrase (use empty options array)

IMPORTANT: Always include the "options" field in your response. Use an array of strings for multiple-choice questions, and an empty array [] for true-false and fill-blank questions.

Ensure all questions are directly related to the slide content and transcript provided.

\end{lstlisting}

\subsection{Highlight Generation}~\label{highlight-generation}
\begin{lstlisting}[label={lst:particles_json}]
I am going to present a slide of a lecture presentation and the transcript from the professor explaining this slide. Identify all visual and textual content blocks on the slide (titles, text, figures, charts). Ignore the visual components out of the lecture slide. Then, decide which part of the transcript is most relavant to the content block. It is ok if there is no relavant transcript.
Output the a list of bounding boxes with the most relavant transcript.
Example:
    [
        {
            "box_2d": [100, 200, 300, 400],
            "relavant_transcript": "relavant transcript, or empty"
        },
        ...
    ]

Transcript: ${slideTranscript}
Only output pure JSON
\end{lstlisting}

\subsection{On-demand Clarification}~\label{on-demand-clarification}
\begin{lstlisting}[label={lst:on-demand}]
# Persona:
You are an AI lecturer.
You are friendly, engaging, and enthusiastic.

# Context:
You are teaching the **${currentVideoName || null}** course.
                          
## Entire Lecture Summary:
${summaryText || null}.

## Current Slide Content:
${currentSlideContent || null}.
                          
# Instruction:
You need to answer students' questions. Keep answer short within 50 words and 3 sentences. Make sure to aswer in smooth, natural-sounding speech transcript. Remove all markdown and LaTeX formatting, and make it as plain text suitable for reading aloud. Symbols like backslashes, parentheses, brackets, and dollar signs should be omitted or read naturally. Ensure the meaning stays accurate, but the phrasing should flow like natural speech. Exclude citation links and references.
\end{lstlisting}

\subsection{Enhanced Visual Search Word Extraction}~\label{enhanced-visual-search-word-extraction}
\begin{lstlisting}[label={lst:enhanced-visual}]
You are a concise OCR+vision tagger. If the image includes text, do OCR and return the most informative keywords or short phrases extracted from that text (strip stopwords, keep domain terms). If little/no text, describe the visual content and return the keywords that would help find better explanatory images. Output ONLY strict JSON: {"keywords":""}, no extra text.

\end{lstlisting}

\subsection{Adaptive Break}~\label{adaptive-break}
\begin{lstlisting}[label={lst:adaptive-break}]
# Persona:
You are an AI lecturer who is taking a break with students.
You are friendly, engaging, and enthusiastic.
You love to tell interesting stories that are both entertaining and somewhat educational.

# Context:
You were teaching the **${currentVideoName}** course.
The students are taking a ${breakDuration}-minute break to refresh their minds.
      
## Lecture Context:
${summaryText || ''}.
${currentSlideContent || ''}.
      
# Instruction:
Tell an interesting story that:
1. Combine with the user's interests about ${userInterests}
2. Is somewhat related to the subject matter but in a fun way
3. Is engaging and entertaining
4. Takes exactly ${breakDuration} minutes to tell (approximately ${breakDuration * 150} words, assuming average speaking speed)
5. Makes students smile and feel refreshed
6. Helps them want to return to learning afterward
      
Keep the story natural and conversational. Avoid technical jargon unless using it in a funny way.
Make sure the story length matches the break duration by adjusting the content accordingly.
\end{lstlisting}

\section{Custom Questionnaire}~\label{section:custom-questionnaire}

Participants rated each item on a 7-point Likert scale (1 = strongly disagree, 7 = strongly agree).

\begin{table}[h]
\centering
\begin{tabular}{p{0.22\linewidth} p{0.70\linewidth}}
\toprule
\textbf{Construct} & \textbf{Question} \\
\midrule
Satisfaction & Overall, I am satisfied with my learning experience using this system. \\
Understanding & This system helped me understand the lecture content better. \\
Engagement & I felt engaged while learning with this system. \\
Review & The system helped me review the lecture content effectively. \\
Focus & The system helped me maintain focus during learning. \\
Adoption & I would be willing to adopt or continue using this system for future learning. \\
\bottomrule
\end{tabular}
\caption{Custom questionnaire items used in the user study.}
\label{tab:custom-questionnaire}
\end{table}

%% file: references.bib
@article{de2018video,
  title={Video-based interactive storytelling using real-time video compositing techniques},
  author={De Lima, Edirlei Soares and Feij{\'o}, Bruno and Furtado, Antonio L},
  journal={Multimedia tools and Applications},
  volume={77},
  pages={2333--2357},
  year={2018},
  publisher={Springer}
}

@article{brooke1996sus,
  title={SUS-A quick and dirty usability scale},
  author={Brooke, John and others},
  journal={Usability evaluation in industry},
  volume={189},
  number={194},
  pages={4--7},
  year={1996},
  publisher={London, England}
}

@article{hart1986nasa,
  title={NASA Task Load Index (TLX): Paper and Pencil Package-Volume 1.0},
  author={Hart, Sandra G},
  year={1986}
}

@article{pataranutaporn2021ai,
  title={AI-generated characters for supporting personalized learning and well-being},
  author={Pataranutaporn, Pat and Danry, Valdemar and Leong, Joanne and Punpongsanon, Parinya and Novy, Dan and Maes, Pattie and Sra, Misha},
  journal={Nature Machine Intelligence},
  volume={3},
  number={12},
  pages={1013--1022},
  year={2021},
  publisher={Nature Publishing Group UK London}
}

@inproceedings{yang2024aqua,
  title={AQuA: Automated Question-Answering in Software Tutorial Videos with Visual Anchors},
  author={Yang, Saelyne and Vermeulen, Jo and Fitzmaurice, George and Matejka, Justin},
  booktitle={Proceedings of the CHI Conference on Human Factors in Computing Systems},
  pages={1--19},
  year={2024}
}

@article{bahroun2023transforming,
  title={Transforming education: A comprehensive review of generative artificial intelligence in educational settings through bibliometric and content analysis},
  author={Bahroun, Zied and Anane, Chiraz and Ahmed, Vian and Zacca, Andrew},
  journal={Sustainability},
  volume={15},
  number={17},
  pages={12983},
  year={2023},
  publisher={MDPI}
}

@article{graesser2005autotutor,
  title={AutoTutor: An intelligent tutoring system with mixed-initiative dialogue},
  author={Graesser, Arthur C and Chipman, Patrick and Haynes, Brian C and Olney, Andrew},
  journal={IEEE Transactions on Education},
  volume={48},
  number={4},
  pages={612--618},
  year={2005},
  publisher={IEEE}
}

@article{pesovski2024generative,
  title={Generative ai for customizable learning experiences},
  author={Pesovski, Ivica and Santos, Ricardo and Henriques, Roberto and Trajkovik, Vladimir},
  journal={Sustainability},
  volume={16},
  number={7},
  pages={3034},
  year={2024},
  publisher={MDPI}
}

@inproceedings{peng2023storyfier,
  title={Storyfier: Exploring vocabulary learning support with text generation models},
  author={Peng, Zhenhui and Wang, Xingbo and Han, Qiushi and Zhu, Junkai and Ma, Xiaojuan and Qu, Huamin},
  booktitle={Proceedings of the 36th Annual ACM Symposium on User Interface Software and Technology},
  pages={1--16},
  year={2023}
}

@inproceedings{panchal2024lingocomics,
  title={LingoComics: Co-Authoring Comic Style AI-Empowered Stories for Language Learning Immersion with Story Designer},
  author={Panchal, Deval and Collins, Christopher and Shimabukuro, Mariana},
  booktitle={Adjunct Proceedings of the 37th Annual ACM Symposium on User Interface Software and Technology},
  pages={1--3},
  year={2024}
}

@inproceedings{kim2015rimes,
  title={RIMES: Embedding interactive multimedia exercises in lecture videos},
  author={Kim, Juho and Glassman, Elena L and Monroy-Hern{\'a}ndez, Andr{\'e}s and Morris, Meredith Ringel},
  booktitle={Proceedings of the 33rd annual ACM conference on human factors in computing systems},
  pages={1535--1544},
  year={2015}
}

@inproceedings{lim2024potential,
  title={The Potential of Learning With AI-Generated Pedagogical Agents in Instructional Videos},
  author={Lim, Jullia},
  booktitle={Extended Abstracts of the CHI Conference on Human Factors in Computing Systems},
  pages={1--6},
  year={2024}
}

@inproceedings{fan2024lessonplanner,
  title={LessonPlanner: Assisting novice teachers to prepare pedagogy-driven lesson plans with large language models},
  author={Fan, Haoxiang and Chen, Guanzheng and Wang, Xingbo and Peng, Zhenhui},
  booktitle={Proceedings of the 37th Annual ACM Symposium on User Interface Software and Technology},
  pages={1--20},
  year={2024}
}

@inproceedings{khokhar2022modifying,
  title={Modifying pedagogical agent spatial guidance sequences to respond to eye-tracked student gaze in VR},
  author={Khokhar, Adil and Borst, Christoph},
  booktitle={Proceedings of the 2022 ACM Symposium on Spatial User Interaction},
  pages={1--12},
  year={2022}
}

@inproceedings{kazemitabaar2024codeaid,
  title={Codeaid: Evaluating a classroom deployment of an llm-based programming assistant that balances student and educator needs},
  author={Kazemitabaar, Majeed and Ye, Runlong and Wang, Xiaoning and Henley, Austin Zachary and Denny, Paul and Craig, Michelle and Grossman, Tovi},
  booktitle={Proceedings of the CHI Conference on Human Factors in Computing Systems},
  pages={1--20},
  year={2024}
}

@inproceedings{zeghouani2024examining,
  title={Examining the Feasibility of AI-Generated Questions in Educational Settings},
  author={Zeghouani, Omar and Ali, Zawar and Simson van Dijkhuizen, William and Hong, Jia Wei and Clos, Jeremie},
  booktitle={Proceedings of the Second International Symposium on Trustworthy Autonomous Systems},
  pages={1--6},
  year={2024}
}

@inproceedings{weerakoon2024enhancing,
  title={Enhancing Pedagogy with Generative AI: Video Production from Course Descriptions},
  author={Weerakoon, Oshani and Lepp{\"a}nen, Ville and M{\"a}kil{\"a}, Tuomas},
  booktitle={Proceedings of the International Conference on Computer Systems and Technologies 2024},
  pages={249--255},
  year={2024}
}

@inproceedings{abbas2024chemgenx,
  title={ChemGenX: AI in the Chemistry Classroom},
  author={Abbas, Touqeer and Javed, Umair and Mehmood, Faisal and Raza, Muneeb and Li, Hui},
  booktitle={Proceedings of the 2024 International Symposium on Artificial Intelligence for Education},
  pages={224--230},
  year={2024}
}

@inproceedings{almadhoob2024quizwiz,
  title={QuizWiz: Integrating Generative Artificial Intelligence in an Online Study Tool},
  author={Almadhoob, Ali Husain and Saleh, Akbar Sayed Kadhem and Akbar, Fatema},
  booktitle={Proceedings of the 2024 7th International Conference on Big Data and Education},
  pages={87--96},
  year={2024}
}

@inproceedings{rajala2023call,
  title={$\backslash$" Call me Kiran$\backslash$"--ChatGPT as a Tutoring Chatbot in a Computer Science Course},
  author={Rajala, Jaakko and Hukkanen, Jenni and Hartikainen, Maria and Niemel{\"a}, Pia},
  booktitle={Proceedings of the 26th International Academic Mindtrek Conference},
  pages={83--94},
  year={2023}
}

@inproceedings{chen2024gptutor,
  title={GPTutor: Great personalized tutor with large language models for personalized learning content generation},
  author={Chen, Eason and Lee, Jia-En and Lin, Jionghao and Koedinger, Kenneth},
  booktitle={Proceedings of the Eleventh ACM Conference on Learning@ Scale},
  pages={539--541},
  year={2024}
}

@inproceedings{abolnejadian2024leveraging,
  title={Leveraging ChatGPT for Adaptive Learning through Personalized Prompt-based Instruction: A CS1 Education Case Study},
  author={Abolnejadian, Mohammad and Alipour, Sharareh and Taeb, Kamyar},
  booktitle={Extended Abstracts of the CHI Conference on Human Factors in Computing Systems},
  pages={1--8},
  year={2024}
}

@inproceedings{calo2024towards,
  title={Towards educator-driven tutor authoring: generative AI approaches for creating intelligent tutor interfaces},
  author={Calo, Tommaso and Maclellan, Christopher},
  booktitle={Proceedings of the Eleventh ACM Conference on Learning@ Scale},
  pages={305--309},
  year={2024}
}

@inproceedings{han2024teachers,
  title={Teachers, Parents, and Students' perspectives on Integrating Generative AI into Elementary Literacy Education},
  author={Han, Ariel and Zhou, Xiaofei and Cai, Zhenyao and Han, Shenshen and Ko, Richard and Corrigan, Seth and Peppler, Kylie A},
  booktitle={Proceedings of the CHI Conference on Human Factors in Computing Systems},
  pages={1--17},
  year={2024}
}

@article{xu2024recorded,
  title={From recorded to AI-generated instructional videos: A comparison of learning performance and experience},
  author={Xu, Tao and Liu, Yuan and Jin, Yaru and Qu, Yueyao and Bai, Jie and Zhang, Wenlan and Zhou, Yun},
  journal={British Journal of Educational Technology},
  year={2024},
  publisher={Wiley Online Library}
}

@inproceedings{dao2021ai,
  title={Ai-powered moocs: Video lecture generation},
  author={Dao, Xuan-Quy and Le, Ngoc-Bich and Nguyen, Thi-My-Thanh},
  booktitle={Proceedings of the 2021 3rd International Conference on Image, Video and Signal Processing},
  pages={95--102},
  year={2021}
}

@article{zheng2023self,
  title={The Self 2.0: How AI-Enhanced Self-Clones Transform Self-Perception and Improve Presentation Skills},
  author={Zheng, Qingxiao and Huang, Yun},
  journal={arXiv preprint arXiv:2310.15112},
  year={2023}
}

@article{seo2021impact,
  title={The impact of artificial intelligence on learner--instructor interaction in online learning},
  author={Seo, Kyoungwon and Tang, Joice and Roll, Ido and Fels, Sidney and Yoon, Dongwook},
  journal={International journal of educational technology in higher education},
  volume={18},
  pages={1--23},
  year={2021},
  publisher={Springer}
}

@article{kim2020my,
  title={My teacher is a machine: Understanding students’ perceptions of AI teaching assistants in online education},
  author={Kim, Jihyun and Merrill, Kelly and Xu, Kun and Sellnow, Deanna D},
  journal={International Journal of Human--Computer Interaction},
  volume={36},
  number={20},
  pages={1902--1911},
  year={2020},
  publisher={Taylor \& Francis}
}

@article{wang2025generative,
  title={Generative Co-Learners: Enhancing Cognitive and Social Presence of Students in Asynchronous Learning with Generative AI},
  author={Wang, Tianjia and Wu, Tong and Liu, Huayi and Brown, Chris and Chen, Yan},
  journal={Proceedings of the ACM on Human-Computer Interaction},
  volume={9},
  number={1},
  pages={1--24},
  year={2025},
  publisher={ACM New York, NY, USA}
}

@article{lee2024teachers,
  title={Teachers' and students' perceptions of AI-generated concept explanations: Implications for integrating generative AI in computer science education},
  author={Lee, Soohwan and Song, Ki-Sang},
  journal={Computers and Education: Artificial Intelligence},
  volume={7},
  pages={100283},
  year={2024},
  publisher={Elsevier}
}

@inproceedings{leong2024putting,
  title={Putting Things into Context: Generative AI-Enabled Context Personalization for Vocabulary Learning Improves Learning Motivation},
  author={Leong, Joanne and Pataranutaporn, Pat and Danry, Valdemar and Perteneder, Florian and Mao, Yaoli and Maes, Pattie},
  booktitle={Proceedings of the CHI Conference on Human Factors in Computing Systems},
  pages={1--15},
  year={2024}
}

@inproceedings{pataranutaporn2023living,
  title={Living memories: AI-generated characters as digital mementos},
  author={Pataranutaporn, Pat and Danry, Valdemar and Blanchard, Lancelot and Thakral, Lavanay and Ohsugi, Naoki and Maes, Pattie and Sra, Misha},
  booktitle={Proceedings of the 28th International Conference on Intelligent User Interfaces},
  pages={889--901},
  year={2023}
}

@inproceedings{pataranutaporn2022ai,
  title={AI-generated virtual instructors based on liked or admired people can improve motivation and foster positive emotions for learning},
  author={Pataranutaporn, Pat and Leong, Joanne and Danry, Valdemar and Lawson, Alyssa P and Maes, Pattie and Sra, Misha},
  booktitle={2022 IEEE Frontiers in Education Conference (FIE)},
  pages={1--9},
  year={2022},
  organization={IEEE}
}

@inproceedings{liu2024classmeta,
  title={ClassMeta: Designing Interactive Virtual Classmate to Promote VR Classroom Participation},
  author={Liu, Ziyi and Zhu, Zhengzhe and Zhu, Lijun and Jiang, Enze and Hu, Xiyun and Peppler, Kylie A and Ramani, Karthik},
  booktitle={Proceedings of the CHI Conference on Human Factors in Computing Systems},
  pages={1--17},
  year={2024}
}

@article{johnson2000animated,
  title={Animated pedagogical agents: Face-to-face interaction in interactive learning environments},
  author={Johnson, W Lewis and Rickel, Jeff W and Lester, James C and others},
  journal={International Journal of Artificial intelligence in education},
  volume={11},
  number={1},
  pages={47--78},
  year={2000},
  publisher={Citeseer}
}

@article{quah2022systematic,
  title={A systematic literature review on digital storytelling authoring tool in education: January 2010 to January 2020},
  author={Quah, Chia Yi and Ng, Kher Hui},
  journal={International Journal of Human--Computer Interaction},
  volume={38},
  number={9},
  pages={851--867},
  year={2022},
  publisher={Taylor \& Francis}
}

@inproceedings{truong2021automatic,
  title={Automatic generation of two-level hierarchical tutorials from instructional makeup videos},
  author={Truong, Anh and Chi, Peggy and Salesin, David and Essa, Irfan and Agrawala, Maneesh},
  booktitle={Proceedings of the 2021 CHI Conference on Human Factors in Computing Systems},
  pages={1--16},
  year={2021}
}

@inproceedings{kim2023papeos,
  title={Papeos: Augmenting Research Papers with Talk Videos},
  author={Kim, Tae Soo and Latzke, Matt and Bragg, Jonathan and Zhang, Amy X and Chang, Joseph Chee},
  booktitle={Proceedings of the 36th Annual ACM Symposium on User Interface Software and Technology},
  pages={1--19},
  year={2023}
}

@inproceedings{clarke2020reactive,
  title={Reactive video: adaptive video playback based on user motion for supporting physical activity},
  author={Clarke, Christopher and Cavdir, Doga and Chiu, Patrick and Denoue, Laurent and Kimber, Don},
  booktitle={Proceedings of the 33rd Annual ACM Symposium on User Interface Software and Technology},
  pages={196--208},
  year={2020}
}

@inproceedings{pavel2015sceneskim,
  title={Sceneskim: Searching and browsing movies using synchronized captions, scripts and plot summaries},
  author={Pavel, Amy and Goldman, Dan B and Hartmann, Bj{\"o}rn and Agrawala, Maneesh},
  booktitle={Proceedings of the 28th Annual ACM Symposium on User Interface Software \& Technology},
  pages={181--190},
  year={2015}
}

@inproceedings{murakami2024swapvid,
  title={SwapVid: Integrating Video Viewing and Document Exploration with Direct Manipulation},
  author={Murakami, Taichi and Fujita, Kazuyuki and Hara, Kotaro and Takashima, Kazuki and Kitamura, Yoshifumi},
  booktitle={Proceedings of the CHI Conference on Human Factors in Computing Systems},
  pages={1--13},
  year={2024}
}

@article{jin2024teachtune,
  title={TeachTune: Reviewing Pedagogical Agents Against Diverse Student Profiles with Simulated Students},
  author={Jin, Hyoungwook and Yoo, Minju and Park, Jeongeon and Lee, Yokyung and Wang, Xu and Kim, Juho},
  journal={arXiv preprint arXiv:2410.04078},
  year={2024}
}

@article{fang2024edulive,
  title={EduLive: Re-Creating Cues for Instructor-Learners Interaction in Educational Live Streams with Learners' Transcript-Based Annotations},
  author={Fang, Jingchao and Park, Jeongeon and Kim, Juho and Wang, Hao-Chuan},
  journal={Proceedings of the ACM on Human-Computer Interaction},
  volume={8},
  number={CSCW2},
  pages={1--33},
  year={2024},
  publisher={ACM New York, NY, USA}
}

@article{blau2021writing,
  title={Writing private and shared annotations and lurking in Annoto hyper-video in academia: Insights from learning analytics, content analysis, and interviews with lecturers and students},
  author={Blau, Ina and Shamir-Inbal, Tamar},
  journal={Educational Technology Research and Development},
  volume={69},
  number={2},
  pages={763--786},
  year={2021},
  publisher={Springer}
}

@article{mirriahi2016uncovering,
  title={Uncovering student learning profiles with a video annotation tool: reflective learning with and without instructional norms},
  author={Mirriahi, Negin and Liaqat, Daniyal and Dawson, Shane and Ga{\v{s}}evi{\'c}, Dragan},
  journal={Educational Technology Research and Development},
  volume={64},
  pages={1083--1106},
  year={2016},
  publisher={Springer}
}

@article{lu2021streamsketch,
  title={StreamSketch: Exploring multi-modal interactions in creative live streams},
  author={Lu, Zhicong and Kazi, Rubaiat Habib and Wei, Li-yi and Dontcheva, Mira and Karahalios, Karrie},
  journal={Proceedings of the ACM on Human-Computer Interaction},
  volume={5},
  number={CSCW1},
  pages={1--26},
  year={2021},
  publisher={ACM New York, NY, USA}
}

@article{chen2017watching,
  title={Watching a movie alone yet together: understanding reasons for watching Danmaku videos},
  author={Chen, Yue and Gao, Qin and Rau, Pei-Luen Patrick},
  journal={International Journal of Human--Computer Interaction},
  volume={33},
  number={9},
  pages={731--743},
  year={2017},
  publisher={Taylor \& Francis}
}

@article{mir2021investigation,
  title={Investigation of students’ satisfaction about H5P interactive video on moodle for online learning},
  author={Mir, Kamran and Iqbal, Muhammad Zafar and Shams, Jahan Ara},
  journal={International Journal of Distance Education and E-Learning},
  volume={7},
  number={1},
  pages={71--82},
  year={2021}
}

@inproceedings{kurzhals2020view,
  title={A view on the viewer: Gaze-adaptive captions for videos},
  author={Kurzhals, Kuno and G{\"o}bel, Fabian and Angerbauer, Katrin and Sedlmair, Michael and Raubal, Martin},
  booktitle={Proceedings of the 2020 CHI Conference on Human Factors in Computing Systems},
  pages={1--12},
  year={2020}
}

@inproceedings{shin2020body,
  title={Body follows eye: Unobtrusive posture manipulation through a dynamic content position in virtual reality},
  author={Shin, Joon Gi and Kim, Doheon and So, Chaehan and Saakes, Daniel},
  booktitle={Proceedings of the 2020 CHI Conference on Human Factors in Computing Systems},
  pages={1--14},
  year={2020}
}

@article{lin2018exploratory,
  title={An exploratory study: using Danmaku in online video-based lectures},
  author={Lin, Xi and Huang, Mingyu and Cordie, Leslie},
  journal={Educational Media International},
  volume={55},
  number={3},
  pages={273--286},
  year={2018},
  publisher={Taylor \& Francis}
}

@inproceedings{riegler2014videojot,
  title={Videojot: A multifunctional video annotation tool},
  author={Riegler, Michael and Lux, Mathias and Charvillat, Vincent and Carlier, Axel and Vliegendhart, Raynor and Larson, Martha},
  booktitle={Proceedings of international conference on multimedia retrieval},
  pages={534--537},
  year={2014}
}

@inproceedings{wu2018danmaku,
  title={Danmaku vs. forum comments: understanding user participation and knowledge sharing in online videos},
  author={Wu, Qunfang and Sang, Yisi and Zhang, Shan and Huang, Yun},
  booktitle={Proceedings of the 2018 ACM International Conference on Supporting Group Work},
  pages={209--218},
  year={2018}
}

@incollection{rama2022enhanced,
  title={Enhanced learning outcomes by interactive video content—H5P in Moodle LMS},
  author={Rama Devi, S and Subetha, Thamaraiselvi and Aruna Rao, Sudha Laxmi and Morampudi, Mahesh Kumar},
  booktitle={Inventive Systems and Control: Proceedings of ICISC 2022},
  pages={189--203},
  year={2022},
  publisher={Springer}
}

@article{elnahla2020black,
  title={Black Mirror: Bandersnatch and how Netflix manipulates us, the new gods},
  author={Elnahla, Nada},
  journal={Consumption Markets \& Culture},
  volume={23},
  number={5},
  pages={506--511},
  year={2020},
  publisher={Taylor \& Francis}
}

@article{leake2017computational,
  title={Computational video editing for dialogue-driven scenes.},
  author={Leake, Mackenzie and Davis, Abe and Truong, Anh and Agrawala, Maneesh},
  journal={ACM Trans. Graph.},
  volume={36},
  number={4},
  pages={130--1},
  year={2017}
}

@article{zhou2019magic,
  title={The magic of danmaku: A social interaction perspective of gift sending on live streaming platforms},
  author={Zhou, Jilei and Zhou, Jing and Ding, Ying and Wang, Hansheng},
  journal={Electronic Commerce Research and Applications},
  volume={34},
  pages={100815},
  year={2019},
  publisher={Elsevier}
}

@inproceedings{wang2024lave,
  title={LAVE: LLM-Powered Agent Assistance and Language Augmentation for Video Editing},
  author={Wang, Bryan and Li, Yuliang and Lv, Zhaoyang and Xia, Haijun and Xu, Yan and Sodhi, Raj},
  booktitle={Proceedings of the 29th International Conference on Intelligent User Interfaces},
  pages={699--714},
  year={2024}
}

@misc{netflix2018bandersnatch,
  author        = "Netflix",
  year          = "2018",
  title         = "Black Mirror: Bandersnatch",
  howpublished  = "An interactive choose-your-own-adventure video experience, allowing viewers to select narrative branches",
  url           = "https://www.netflix.com/watch/80238138",
  lastaccessed  = "Dec 18, 2024",
}

@inproceedings{chang2021rubyslippers,
  title={Rubyslippers: Supporting content-based voice navigation for how-to videos},
  author={Chang, Minsuk and Huh, Mina and Kim, Juho},
  booktitle={Proceedings of the 2021 CHI conference on human factors in computing systems},
  pages={1--14},
  year={2021}
}

@article{braun2006using,
  title={Using thematic analysis in psychology},
  author={Braun, Virginia and Clarke, Victoria},
  journal={Qualitative research in psychology},
  volume={3},
  number={2},
  pages={77--101},
  year={2006},
  publisher={Taylor \& Francis}
}

@inproceedings{kim2021hyperbutton,
  title={HyperButton: In-video Question Answering via Interactive Buttons and Hyperlinks},
  author={Kim, Jeongyeon and Park, Junyong and Lu, I Hao},
  booktitle={Proceedings of the Asian CHI Symposium 2021},
  pages={48--52},
  year={2021}
}

@inproceedings{chi2012mixt,
  author    = {Chi, Pei{-}Yu (Peggy) and Ahn, Sally and Ren, Amanda and Dontcheva, Mira and Li, Wilmot and Hartmann, Bj{\"o}rn},
  title     = {MixT: Automatic Generation of Step-by-Step Mixed Media Tutorials},
  booktitle = {Proceedings of the 25th Annual ACM Symposium on User Interface Software and Technology},
  series    = {UIST '12},
  pages     = {93--102},
  year      = {2012},
  publisher = {Association for Computing Machinery},
  address   = {New York, NY, USA}
}

@inproceedings{chi2021howtocut,
  author    = {Chi, Peggy and Frey, Nathan and Panovich, Katrina and Essa, Irfan},
  title     = {Automatic Instructional Video Creation from a Markdown-Formatted Tutorial},
  booktitle = {Proceedings of the 34th Annual ACM Symposium on User Interface Software and Technology},
  series    = {UIST '21},
  year      = {2021},
  publisher = {Association for Computing Machinery},
  address   = {New York, NY, USA},
  doi       = {10.1145/3472749.3474778}
}

@inproceedings{chi2022doc2video,
  author    = {Chi, Peggy and Dong, Tao and Colonna, Brian R. and Frueh, Christian and Kwatra, Vivek and Essa, Irfan},
  title     = {Synthesis-Assisted Video Prototyping From a Document},
  booktitle = {Proceedings of the 35th Annual ACM Symposium on User Interface Software and Technology},
  series    = {UIST '22},
  year      = {2022},
  publisher = {Association for Computing Machinery},
  address   = {New York, NY, USA}
}

@inproceedings{chi2020url2video,
  author    = {Chi, Peggy and Sun, Zheng and Panovich, Katrina and Essa, Irfan},
  title     = {URL2Video: Automatic Video Creation From a Web Page},
  booktitle = {Proceedings of the 33rd Annual ACM Symposium on User Interface Software and Technology},
  series    = {UIST '20},
  pages     = {279--292},
  year      = {2020},
  publisher = {Association for Computing Machinery},
  address   = {New York, NY, USA}
}

@inproceedings{zhong2021helpviz,
  author    = {Zhong, Mingyuan and Li, Gang and Chi, Peggy and Li, Yang},
  title     = {HelpViz: Automatic Generation of Contextual Visual Mobile Tutorials from Text-Based Instructions},
  booktitle = {Proceedings of the 34th Annual ACM Symposium on User Interface Software and Technology},
  series    = {UIST '21},
  year      = {2021},
  publisher = {Association for Computing Machinery},
  address   = {New York, NY, USA},
  doi       = {10.1145/3472749.3474812}
}

@inproceedings{cao_videosticker,
  author    = {Cao, Yining and Subramonyam, Hariharan and Adar, Eytan},
  title     = {VideoSticker: A Tool for Active Viewing and Visual Note-taking from Videos},
  booktitle = {Proceedings of the 27th International Conference on Intelligent User Interfaces},
  series    = {IUI '22},
  year      = {2022},
  pages     = {672--690},
  publisher = {Association for Computing Machinery},
  address   = {New York, NY, USA},
  doi       = {10.1145/3490099.3511132}
}
